\pdfoutput=1
\documentclass[review, fleqn, authoryear]{elsarticle}

\usepackage{algorithm, amsmath, amssymb, bbold, bm, graphicx, here, lineno}
\usepackage{algpseudocode}
\usepackage[colorlinks=true]{hyperref}
\newcommand{\Equref}[1]{Eq.~(\ref{#1})}
\newcommand{\Figref}[1]{Fig.~\ref{#1}}
\newcommand{\argmax}{\mathop\mathrm{argmax}\limits}
\newcommand{\argmin}{\mathop\mathrm{argmin}\limits}
\modulolinenumbers[5]
\journal{Journal of \LaTeX\ Templates}

\usepackage{natbib}
\graphicspath{{./figures/}, {./figures/real_8068/}}

\begin{document}
\begin{frontmatter}
\title{Estimation of neural connections from partially observed neural spikes}

\author[1]{Taishi Iwasaki}
\author[2]{Hideitsu Hino}
\author[3]{Masami Tatsuno}
\author[4]{Shotaro Akaho}
\author[1]{Noboru Murata}
\address[1]{Department of Electrical Engineering and Bioscience, Waseda University, Okubo 3-4-1, Shinjuku-ku, Tokyo 169-0072, Japan}
\address[2]{Department of Computer Science, University of Tsukuba, Tennoudai 1-1-1, Tsukuba, Ibaraki 305-8573, Japan}
\address[3]{Department of Neuroscience, University of Lethbridge, 4401 University Drive, Lethbridge, Alberta T1K 6T5, Canada}
\address[4]{Mathematical Neuroinformatics Group, National Institute of Advanced Industrial Science and Technology, Umezono 1-1-1 Tsukuba, Ibaraki 305-8568, Japan}

\begin{abstract}
Plasticity is one of the most important properties of the nervous system, which enables animals to adjust their behavior to the ever-changing external environment.
Changes in synaptic efficacy between neurons constitute one of the major mechanisms of plasticity. Therefore, estimation of neural connections is crucial for investigating information processing in the brain.
Although many analysis methods have been proposed for this purpose, most of them suffer from one or all the following mathematical difficulties: (1) only partially observed neural activity is available; (2) correlations can include both direct and indirect pseudo-interactions; and (3) biological evidence that a neuron typically has only one type of connection (excitatory or inhibitory) should be considered.
To overcome these difficulties, a novel probabilistic framework for estimating neural connections from partially observed spikes is proposed in this paper.
First, based on the property of a sum of random variables, the proposed method estimates the influence of unobserved neurons on observed neurons and extracts only the correlations among observed neurons.
Second, the relationship between pseudo-correlations and target connections is modeled by neural propagation in a multiplicative manner.
Third, a novel information-theoretic framework is proposed for estimating neuron types.
The proposed method was validated using spike data generated by artificial neural networks.
In addition, it was applied to multi-unit data recorded from the CA1 area of a rat's hippocampus. The results confirmed that our estimates are consistent with previous reports.
These findings indicate that the proposed method is useful for extracting crucial interactions in neural signals as well as in other multi-probed point process data.
\end{abstract}


\end{frontmatter}

\linenumbers
\section{Introduction}
One of the most important properties of the brain is its ability to modify its architecture on the basis of experience.
This phenomenon, which is known as plasticity, enables living organisms to flexibly adjust their behavior to the external environment and improve their chances of survival.
Various studies have shown that changes in synaptic connections constitute the primary mechanism of plasticity, in which many types of neurotransmitters and receptors are involved.
Although the detailed processes of how synaptic efficacy is modified are complex, detecting the overall change in neural connections is essential for investigating information processing in the brain.
Recent advancements in experimental technologies, such as multi-electrode recording from a freely behaving animal, enable us to record the activities of a large number of neurons simultaneously for extended periods~\citep{Tatsuno2006methodological}.
After spike sorting of the multi-unit activity (MUA) data, sorted single-unit activity (SUA) data are obtained.
SUA represents the timing of spike occurrence of each neuron, and it can be considered as a point process.
Many studies have investigated the correlational properties of SUA data with the objective of understanding system-level information processing in the brain~\citep{perkel1967neuronal, gerstein1969simultaneously, brown2004multiple, shimazaki2012state, hino2015mmpp, takano2015patchworking}.
Toward this end, methods based on pairwise neuronal correlations, such as cross-correlation~\citep{wilson1994reactivation, bartho2004characterization} and joint peristimulus time histogram~\citep{aertsen1989dynamics, ito2000model} have been widely adopted.
For instance, in the context of memory consolidation, \citet{wilson1994reactivation} estimated functional neural interactions in the CA1 area of the hippocampus by means of cross-correlation functions.
They showed that pairwise correlations induced during a behavior task epoch were sustained during a post-task non-REM sleep epoch, thereby supporting the conjecture that reactivation of behaviorally induced neural activity during sleep (memory reactivation) plays an important role in memory consolidation.
Other approaches based on graph structure estimation methods, such as sparse inverse covariance selection~(SICS), have been also adopted~\citep{banerjee2008model, friedman2008sparse, scheinberg2009sinco}.
SICS assumes that observed data are generated from a Gaussian distribution and estimates a graph structure as its inverse covariance matrix.
Efficient algorithms for SICS have been proposed, and they can estimate the functional connections of a network composed of numerous neurons.
However, the above-mentioned methods suffer from several mathematical difficulties.
First, owing to their focus on pairwise relationships, they cannot capture high-order correlations that might result in pseudo-correlations with pairwise measurements.
Second, they generally provide functional correlations, which lack directional properties as a fundamental feature; therefore, it is difficult to discuss the direction of connections.
Third, they usually do not consider the fact that only a limited number of neurons are recorded in experiments.
Unobserved neurons affect the activity of observed neurons, but existing methods do not include a systematic treatment for interference from unobserved neurons.
Recently, several attempts have been made to overcome the above-mentioned difficulties.
For instance, some studies have adopted the information-theoretic approach to investigate high-order correlations~\citep{nie2012information, nakahara2002information, tatsuno2009information}.
Assuming that spikes are generated from an exponential family of distributions, the method based on information geometry models the probability of coincident multi-neuronal firings, $p_{x_1, x_2, \cdots, x_k}$, by a log-linear model:
\begin{align}
  \ln{p_{x_1,x_2,\cdots, x_k}} = \sum_{i}\theta_i x_i + \sum_{i<j}\theta_{ij} x_ix_j + \cdots + \theta_{12\cdots k}x_1x_2\cdots x_k - \psi,
\end{align}
where $x_i$ is a binary variable representing the spikes of neuron $i$, $\theta$ is a parameter representing neural interactions, $\psi$ is a normalization factor for the integral to be 1, and $k$ is the number of observed neurons, following the same notation as that in the original paper.
Further, $\theta_{ij}$ represents the interaction between neurons $i$ and $j$.
Although this model mitigates the problem of pseudo-correlations, it suffers from two drawbacks. First, its computational cost increases with $k$. Second, it assumes that connections are symmetric.
For the problem of directionality, a method based on Granger causality has been proposed to extract information regarding the direction of connections~\citep{arnold2007temporal, kim2011granger, quinn2011estimating, hu2015copula}.
Suppose that the activities of two neurons, $x_i$ and $x_j$, are observed.
If $x_j$ provides statistically significant improvements of the future values of $x_i$, the directed influence from $x_j$ to $x_i$ is estimated, and it is said that there is Granger causality from $x_j$ to $x_i$.
However, it is difficult to capture higher-order correlations with this approach, which focuses on two spike trains.
To overcome the problems of high-order correlations and directionality simultaneously, \citet{noda2014intrinsic} recently proposed the graph structure estimation method based on the graph Laplacian.
They modeled the correlations between nodes $i$ and $j$, including higher-order ones, by
\begin{align}
	\xi_{ij}
	= c_0
	+ c_{ij}\theta_{ij}
	+ \sum_{k} c_{ij}^{k} \theta_{ik}\theta_{kj}
	+ \sum_{k, j} c_{ij}^{kl} \theta_{ik}\theta_{kl}\theta_{lj}
	+ \cdots
	\nonumber\\
	\qquad {\rm for} \quad i \neq j,
\end{align}
where $\theta_{ij}$ is the connection from $j$ to $i$ and $c$ is a decay coefficient, following the same notation as that in the original paper.
Assuming that the influence deteriorates as it propagates to other neurons, the method models the propagation of influence on the basis of random walk.
However, this method is inadequate for estimating inhibitory connections because the influence can take only positive values.
Finally, regarding the problem of unobserved neurons, to the best of our knowledge, no existing method can explicitly deal with the influence of unobserved neurons.
In this study, we develop a novel mathematical framework that can systematically address the problems of pseudo-correlations, directed connections, and the influence of unobserved neurons.

\section{Problem setting}
In this section, we introduce the notations used and corresponding assumptions.

Suppose that $N$ neurons out of many are observed.
Let $X_i(t)\in\{1, 0\}$ be a random variable representing the state of neuron $i\in\{1, 2, \cdots, N\}$ at time $t$\;(i.e., 1 denotes firing and 0 denotes non-firing).
These neurons' activities are recorded at time $t=1, 2, \cdots, T$ discretely, and the spike data are given by
\begin{align}
	\mathcal{D} =
  \bigl\{X_1(t), X_2(t), \cdots, X_N(t)\bigr\}_{t=1}^T.
\end{align}
Neural connections are represented as a graph structure.
Let $V$ be a set of nodes $\{1, 2, \cdots, N\}$ and $E$ be a set of edges $\bigl\{(i, j), i, j\in V\bigr\}$.
Neural connections are characterized by a graph $(V, E)$, where $V$ corresponds to a set of neurons and $E$ corresponds to a set of synaptic connections.
The graph is also represented by a matrix $W\in\mathbb{R}^{N\times N}$.
Let $w_{ij}$ be element $(i, j)$ of $W$.
Here, $w_{ij}$ represents the strength of a connection from neuron $j$ to neuron $i$.
Connections are classified into two types, namely excitatory and inhibitory connections.
According to neuroscience, a neuron is known to have only one type of connection; thus, neurons are either excitatory or and inhibitory.
Excitatory neurons promote firing of the neurons to which they connect.
On the other hand, inhibitory neurons suppress firing of the connected neurons.
Therefore, $w_{ij}$ is classified as follows:
\begin{align}
	\left\{
	\begin{aligned}
		&w_{ij}>0,&
		\qquad&\mbox{excitatory connection from $j$ to $i$}, \\
		&w_{ij}=0,&
		\qquad&\mbox{no connection from $j$ to $i$},\\
		&w_{ij}<0,&
		\qquad&\mbox{inhibitory connection from $j$ to $i$}.
	\end{aligned}
	\right.
	\label{eq:def_W}
\end{align}
We assume that observed neurons do not have self-connections and that the strength of connection $w_{ij}$ remains unchanged during the observed period $[1, T]$.

\section{Stochastic firing model}
We assume that firing of neuron $i$ at time $t$ is determined by its internal state $U_i(t)$ that corresponds to the membrane potential:
\begin{align}
	\mathrm{Pr}\bigl(X_i(t)=1\bigr)
	= \Phi\bigl(U_i(t)\bigr),
\end{align}
where $\Phi$ is the cumulative distribution function of probability density function $\phi$,
\begin{align}
	\Phi(x) &= \int_{-\infty}^x\phi(z)dz.
	\label{eq:cumfunc}
\end{align}
In this study, for mathematical simplicity, we assume that the probability density function $\phi$ is a Gaussian distribution with mean 0 and variance $\sigma^2$.
The Gaussian distribution function is denoted by $\phi_{\sigma^2}$ and the cumulative distribution function is denoted by $\Phi_{\sigma^2}$:
\begin{align}
  \Phi_{\sigma^2}(x)
  &= \int_{-\infty}^x\phi_{\sigma^2}(z)dz,
	\label{eq:gauss_cumfunc}
	\\
	\phi_{\sigma^2}(z)
	&= \frac{1}{\sqrt{2\pi\sigma^2}}\exp{\Bigl(-\frac{z^2}{2\sigma^2}\Bigr)}.
	\label{eq:gauss_func}
\end{align}
We consider that the internal state $U_i(t)$ is affected by both observed and unobserved neurons:
\begin{align}
	U_i(t)
  = B_i(t) + \sum_{j=1}^{N} w_{ij} X_j(t-S_{ij}),
	\label{eq:inner}
\end{align}
where $S_{ij}$ is a random variable representing the transmission delay from neuron $j$ to neuron $i$ and $B_i(t)$ is a random variable representing the input from unobserved neurons.
In practice, the influence from neuron $j$ to neuron $i$ can be examined using a short time window because the delay $S_{ij}$ is unknown.
Therefore, we introduce a stochastic firing model with a short window:
\begin{align}
	U_i(t)
	= \underbrace{B_i(t)}_{\rm nuisance\;input} + \underbrace{\sum_{j=1}^{N} \lambda_{ij} X_j[t_\Delta]}_{\rm pseudo\mathchar`-correlations},
	\label{eq:inner}
\end{align}
where $[t_\Delta]$ is a short period of time, $[t_{\Delta}]=[t-\Delta, t-\Delta+1, \cdots, t-1](\Delta\in\mathbb{N})$, and $X_j[t_{\Delta}]\in\{1, 0\}$ is a random variable representing the state of neuron $i$ during period $[t_{\Delta}]$\;(i.e., it equals 1 if at least one spike exists during $[t_\Delta]$ and 0 otherwise).
We assume that the length of the time window $\Delta$ is longer than the transmission delays $S_{ij}$ of any pair of observed neurons: $\Delta>S_{ij}, \forall_{(i, j)}$.
The influence from neuron $j$ to neuron $i$ during $[t_\Delta]$ is denoted by $\lambda_{ij}$, including not only a direct influence through connection $w_{ij}$ but also an indirect influence via other neurons.
We refer to such an indirect influence as a {\it pseudo-correlation} and to the coefficient $\lambda_{ij}$ as a {\it pseudo-connection}．

\section{Proposed method}
\begin{figure}[h]
  \centering
  \includegraphics[bb=0 0 981 298, width=\linewidth]{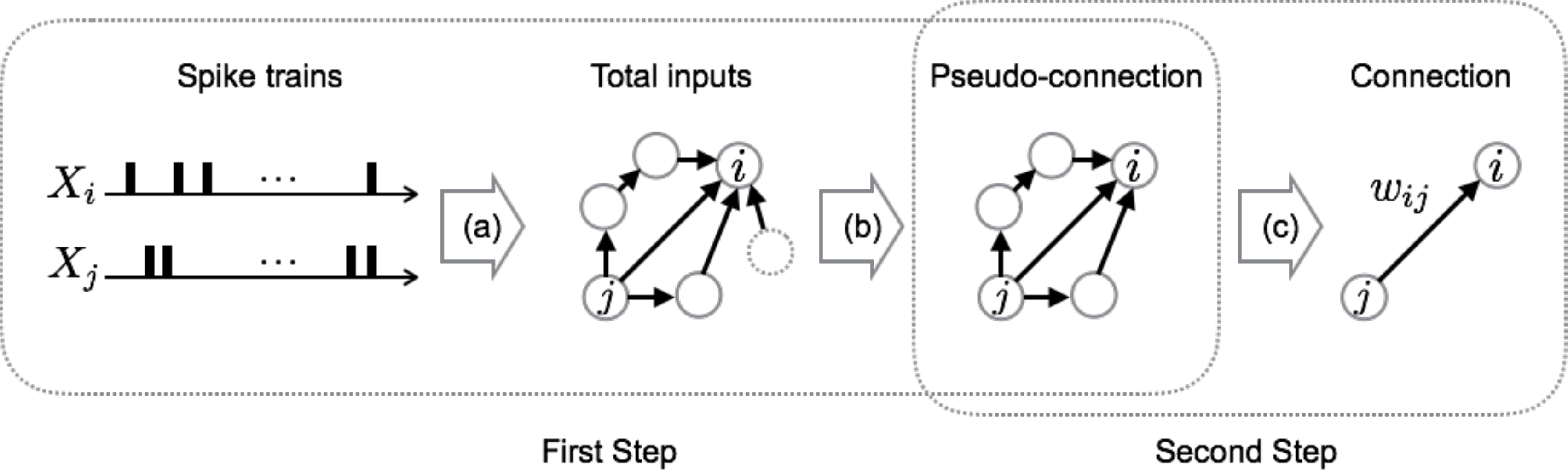}
  \caption{Schematic of the proposed method.
  In the first step ((a) and (b)), pseudo-connection $\lambda_{ij}$ is estimated from spike trains.
  The first procedure (a) is to estimate the total inputs, including the nuisance input, to neuron $i$.
  The second procedure (b) is to remove the nuisance input and estimate pseudo-connection $\lambda_{ij}$. In the second step (c), connection $w_{ij}$ is estimated by decomposing pseudo-connection $\lambda_{ij}$}
  \label{fig:diagram}
\end{figure}

In this section, we first characterize the property of a sum of random variables and discuss how to deal with inputs from unobserved neurons.
Together with the stochastic firing model, we propose a method to estimate pseudo-connection $\lambda_{ij}$ from spike trains.
Then, we propose a method to estimate connection $w_{ij}$ from pseudo-connection $\lambda_{ij}$.
A schematic of the proposed method is shown in \Figref{fig:diagram}.

\subsection{Property of a sum of random variables}
Consider random variables $X$ and $Y$ with probability density functions $f_X$ and $f_Y$, respectively.

\noindent
{\bf Theorem 1.}
	{\sl Let $X$ and $Y$ be independent random variables.
	Then, for any bounded function $g$, we have
	\begin{align}
		{\mathbb E}[g(X+Y)] &= {\mathbb E}\bigl[h\bigl(X+\mathbb{E}[Y]\bigr)\bigr],
		\nonumber\\
		h &= g*f^{-}_{Y},
		\label{eq:Hyv_gene}
	\end{align}
	where $f^{-}_{Y}(x) = f_Y(\mathbb{E}[Y]-x)$.
	}

\noindent
If the function $g$ is a Gaussian density function, we have the following corollary.

\noindent
{\bf Corollary~1.}
	{\sl If the function $g$ is $\phi_{\sigma^2}$ in \Equref{eq:gauss_func} and the probability density function $f_Y$ is a Gaussian distribution with mean $\mathbb{E}[Y]$ and variance $\tau^2$, we have
	\begin{align}
		{\mathbb E}[\phi_{\sigma^2}(X+Y)] = {\mathbb E}\bigl[\phi_{\sigma^2+\tau^2}\bigl(X+\mathbb{E}[Y]\bigr)\bigr].
		\label{eq:Hyv_gene}
	\end{align}
}

\noindent
When $\mathbb{E}[Y]=0$, \Equref{eq:Hyv_gene} is equivalent to  the theorem for Gaussian moments given by \citet{hyvarinen1999gaussian}.
The proof of Theorem~1 and Corollary~1 is given in Appendix~A.

\noindent
{\bf Theorem 2.}
	{\sl Let a function $G$ be the cumulative distribution function of a probability density function $g$.
  If random variables $X$ and $Y$ are independent, we have
	\begin{align}
		{\mathbb E}[G(X+Y)] &= {\mathbb E}\bigl[H\bigl(X+\mathbb{E}[Y]\bigr)\bigr],
		\\
		H(x) &= \int_{-\infty}^{x} (g*f^{-}_{Y})(z) dz,
	\end{align}
	where $f^{-}_{Y}(x) = f_Y(\mathbb{E}[Y]-x)$.
	}

\noindent
We also have the following corollary for a cumulative Gaussian distribution function $\Phi_{\sigma^2}$.

\noindent
{\bf Corollary~2.}
  {\sl If the function $G$ is $\Phi_{\sigma^2}$ in \Equref{eq:gauss_cumfunc}, the random variable $X$ takes a constant value $x$, and $Y\sim\mathcal{N}(\mathbb{E}[Y], \tau^2)$, we have
	\begin{align}
		{\mathbb E}[\Phi_{\sigma^2}(x+Y)] = \Phi_{\sigma^2+\tau^2}\bigl(x+\mathbb{E}[Y]\bigr).
		\label{eq:Hyv_cum}
	\end{align}
	}
\noindent
The proof of Theorem~2 and Corollary~2 can be found in Appendix~A.

\subsection{Removal of nuisance effects}
We introduce a framework for removing the nuisance input and extracting only pseudo-connection $\lambda_{ij}$.
This is achieved by combining the property of a sum of random variables discussed in the previous subsection and the firing probabilities based on a stochastic firing model.
Regarding the influence from neuron $j$ to neuron $i$, we treat two cases separately; one is that neuron $j$ fires and the other is that neuron $j$ does not fire.

First, consider the case in which neuron $j$ fires and then neuron $i$ fires during the period $[t_\Delta]$.
In this case, firing of neuron $i$ is caused by the influence from firing of neuron $j$ and other neurons.
Therefore, the internal state conditioned on $X_j[t_{\Delta}]=1$ is represented by
\begin{align}
	U_{i}\bigl(t\mid X_j[t_{\Delta}]=1\bigr)
	&= B_i(t) + \lambda_{ij}X_j[t_{\Delta}]
		+ \sum_{k \neq j}\lambda_{ik}X_k[t_{\Delta}]
	\nonumber\\
	&= \lambda_{ij} + C_{ij}\bigl(t\mid X_j[t_{\Delta}]=1\bigr),
	\label{eq:inner_long}
\end{align}
where $C_{ij}\bigl(t\mid X_j[t_{\Delta}]=1\bigr)$ is a random variable representing a nuisance input: $C_{ij}\bigl(t\mid X_j[t_{\Delta}]=1\bigr) = B_i(t) + \sum_{k\neq j}\lambda_{ik}X_k[t_{\Delta}]$.
The expectation of neuron $i$'s state $X_i(t)$ conditioned on $X_j[t_{\Delta}]=1$ is obtained as
\begin{align}
	\mathbb{E}\bigl[X_i(t) \mid X_j[t_{\Delta}]=1\bigr]
	&= \mathbb{E}\Bigl[\Phi_{\sigma^2}\Bigl(U_{i}\bigl(t\mid X_j[t_{\Delta}]=1\bigr)\Bigr)\Bigr]
	\nonumber\\
	&= \mathbb{E}\bigl[\Phi_{\sigma^2}\bigl(\lambda_{ij} + C_{ij}(t\mid X_j[t_{\Delta}]=1\bigr)\bigr)\bigr].
	\label{eq:E_given1}
\end{align}
To simplify the formula further, we make the following assumption.

\noindent
{\bf Assumption~1.}
	{\sl In the case that neuron $j$ fires, the nuisance input $C_{ij}\bigl(t\mid X_j[t_{\Delta}]=1\bigr)$ is subject to a Gaussian distribution with mean $\bar{C}_{ij}$ and variance $\tau^2$.
	}
\noindent

\noindent
Under this assumption, we can apply Corollary 2 to the right-hand side of \Equref{eq:E_given1} and obtain
\begin{align}
  \mathbb{E}\Bigl[\Phi_{\sigma^2}\Bigl(\lambda_{ij} + C_{ij}\bigl(t\mid X_j[t_{\Delta}]=1\bigr)\Bigr)\Bigr]
	= \Phi_{\rho^2}\bigl(\lambda_{ij} + \bar{C}_{ij}\bigr),
	\label{eq:E_given1_Hyv}
\end{align}
where $\rho^2 = \sigma^2+\tau^2$.
In addition, the expectation of spike train $X_i(t)$ is equal to its firing probability, i.e.,
\begin{align}
	\mathbb{E}\bigl[X_i(t) \mid X_j[t_{\Delta}]=1\bigr] = \mathrm{Pr}(X_i(t)=1 \mid X_j[t_{\Delta}]=1).
	\label{eq:exp_freq}
\end{align}
Thus, we obtain the following relationship:
\begin{align}
	\Phi_{\rho^2}\bigl(\lambda_{ij} + \bar{C}_{ij}\bigr)
	&= \mathrm{Pr}(X_i(t)=1 \mid X_j[t_{\Delta}]=1)
	\nonumber\\
	\Leftrightarrow
	\lambda_{ij} \!+\! \bar{C}_{ij} &\!=\! \rho\cdot\Phi^{-1}_{1}\Bigl(\mathrm{Pr}(X_i(t)=1 \mid X_j[t_{\Delta}]=1)\Bigr).
	\label{eq:lamb_nonObs}
\end{align}
With an empirical estimate of the conditional probability $\mathrm{Pr}(X_i(t)=1 \mid X_j[t_{\Delta}]=1)$ by using spike trains $X_i$ and $X_j$, we can estimate the sum of pseudo-connection $\lambda_{ij}$ and other influence $\bar{C}_{ij}$.

Second, consider the case in which neuron $j$ does not fire but neuron $i$ fires during the period $[t_\Delta]$.
Firing of neuron $i$ is caused only by the influence from other neurons besides neuron $j$:
\begin{align}
	U_{i}\bigl(t\mid X_j[t_{\Delta}]=0\bigr)
	= C_{ij}\bigl(t\mid X_j[t_{\Delta}]=0\bigr).
\end{align}
As in the case of $X_j[t_{\Delta}]=1$, we make the following assumption.

\noindent
{\bf Assumption~2.}
	{\sl In the case that neuron $j$ does not fire, the nuisance input $C_{ij}(t\mid X_j[t_{\Delta}]=0\bigr)$ is subject to a Gaussian distribution with mean $\bar{C}_{ij}$ and variance $\tau^2$.
	}

\noindent
Note that this is the same distribution as that for the nuisance input $C_{ij}(t\mid X_j[t_{\Delta}]=1\bigr)$ because the effect of the firing of neuron $j$ is negligible compared to the nuisance input from a large number of other neurons.
In other words, the difference between these two nuisance inputs is whether the input includes the influence from neuron $j$.
It is reasonable to assume that one of many neurons' firings does not change a distribution.
Under this assumption, we obtain the following relationship in the same way as \Equref{eq:lamb_nonObs}:
\begin{align}
	\bar{C}_{ij}
	= \rho\cdot\Phi^{-1}_{1}\Bigl(\mathrm{Pr}(X_i(t)=1 \mid X_j[t_{\Delta}]\!=\!0)\Bigr).
	\label{eq:nonObs}
\end{align}
By taking the difference between \Equref{eq:lamb_nonObs} and \Equref{eq:nonObs}, we have
\begin{align}
	\lefteqn{
	\lambda_{ij}
	=\rho\Bigl\{\Phi^{-1}_{1}\Bigl(\mathrm{Pr}(X_i(t)=1 \mid X_j[t_{\Delta}]\!=\!1)\Bigr)
	}\qquad\qquad\quad\nonumber\\
	&- \Phi^{-1}_{1}\Bigl(\mathrm{Pr}(X_i(t)=1 \mid X_j[t_{\Delta}]\!=\!0)\Bigr)
	\Bigr\}.
	\label{eq:est_lambda}
\end{align}
We can exclude the influence from unobserved neurons and extract pseudo-connection $\lambda_{ij}$ up to constant $\rho$, which is defined in \Equref{eq:E_given1_Hyv}.
We note that the firing probabilities in \Equref{eq:est_lambda} are calculated by
\begin{align}
	\mathrm{Pr}(X_i(t)=&1 \mid X_j[t_{\Delta}]\!=\!1)
	\nonumber\\
	&= \frac{1}{F_j}\sum_{t=\Delta+1}^{T} X_i(t \mid X_j[t_\Delta]=1)
	\\
	\mathrm{Pr}(X_i(t)=&1 \mid X_j[t_{\Delta}]\!=\!0)\nonumber\\
	&= \frac{1}{(T-\Delta)-F_j}\sum_{t=\Delta+1}^{T} X_i(t \mid X_j[t_\Delta]=0),
\end{align}
where $F_j = \sum_{t=\Delta+1}^{T} X_j[t_\Delta]$.

\subsection{Decomposition of pseudo-connection to estimate connection}
We propose a method to estimate connection $w_{ij}$ from pseudo-connection $\lambda_{ij}$ obtained by \Equref{eq:est_lambda}.
The key idea is the introduction of virtual propagation probability during the period $[t-\delta, t)$, which is shorter than the period $[t-\Delta, t)$; this enables us to model the relationship between connection $w_{ij}$ and pseudo-connection $\lambda_{ij}$.

\subsubsection{Estimation of virtual propagation probability}
We consider a short period $[t_\delta]=[t-\delta, t)$ in which only a direct influence of neuron $j$ on neuron $i$ is observed.
In this case, the internal state of neuron $i$ in the period $[t_{\delta}]$ is represented by
\begin{align}
	U_{i}\bigl(t\mid X_j[t_{\delta}]=1\bigr)
	= w_{ij} + C'_{ij}(t),
	\label{eq:inner_short}
\end{align}
where $C'_{ij}(t)$ is the input from unobserved neurons at time $t$.
Let $\theta_{ij}$ denote the virtual propagation probability of neuron $i$ after neuron $j$ fires during the period $[t_\delta]$:
\begin{equation}
	\theta_{ij} = \mathrm{Pr}(X_i(t)=1 \mid X_j[t_{\delta}]=1).
	\label{eq:theta_def}
\end{equation}
This probability is equal to the expectation of status $X_i$ conditioned on $X_j[t_{\delta}]=1$.
Therefore, in the same way as \Equref{eq:lamb_nonObs}, \Equref{eq:theta_def} becomes
\begin{equation}
	\theta_{ij}
	= \mathbb{E}\bigl[\Phi_{\sigma^2}(w_{ij}+C'_{ij})\bigr]
	= \Phi_{\rho^2}\bigl(w_{ij}+\mathbb{E}[C'_{ij}]\bigr).
	\label{eq:theta_phi}
\end{equation}
Now, we make the following assumption.

\noindent
{\bf Assumption~3.}
  {\sl Nuisance inputs $C_{ij}(t)$ in the longer period of $[t_{\Delta}]$ and $C'_{ij}(t)$ in the shorter period of $[t_{\delta}]$ are subject to the same Gaussian distribution, with mean $\bar{C}_{ij}$ and variance $\tau^2$.
  }

\noindent
This means that the observed input in $C_{ij}(t)\Bigl(= B_i(t) + (\mbox{observed input})\Bigr)$ is sufficiently smaller than the nuisance input $B_i(t)$.
Because of this assumption, the short period $\delta$ does not need to be specified when detecting the direct transition.
Then, \Equref{eq:theta_phi} becomes
\begin{equation}
	\theta_{ij}
	= \Phi_{\rho^2}(w_{ij}+\bar{C}_{ij}).
	\label{eq:theta_w}
\end{equation}

\subsubsection{Decomposition of pseudo-connection with direct connection}
\begin{figure}
	\centering
	\includegraphics[bb=0 0 699 136, width=\linewidth]{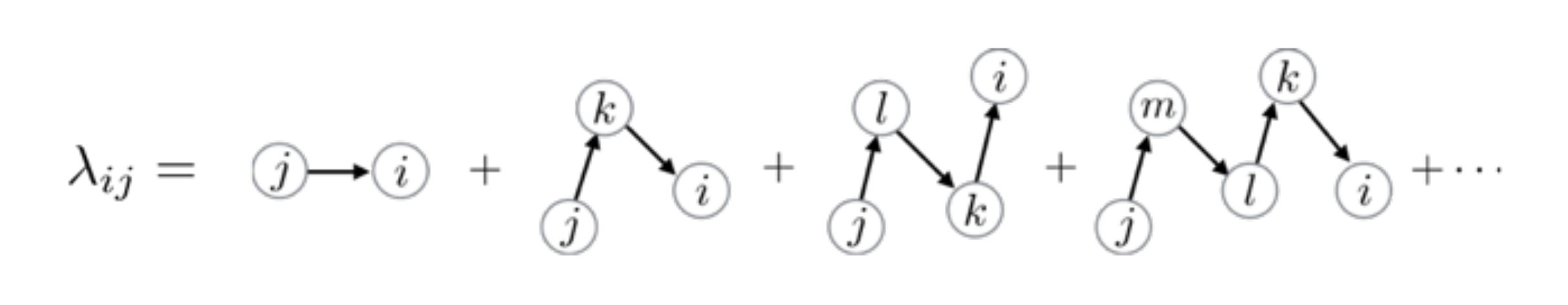}
	\caption{Decomposition of pseudo-connection $\lambda_{ij}$ into multi-step connections}
	\label{fig:pseudo}
\end{figure}

The pseudo-connection $\lambda_{ij}$ is regarded as the averaged influence from neuron $j$ on neuron $i$.
As shown in \Figref{fig:pseudo}, the pseudo-connection $\lambda_{ij}$ includes not only a direct influence $w_{ij}$ but also an indirect influence via other neurons.
We make the following assumption to decompose the pseudo-connection.

\noindent
{\bf Assumption~4.}
{\sl Indirect connections in the pseudo-connection $\lambda_{ij}$ propagate in a multiplicative manner.}

\noindent
The strength of the connection from neuron $j$ to neuron $i$ through neuron $k$ can be regarded as the expectation of the input to neuron $i$ from neuron $k$, which is affected by neuron $j$.
It is represented by $w_{ik}\cdot\mathrm{Pr}(X_k(t)=1\mid X_j[t_{\delta}]=1)$, i.e., $w_{ik}\theta_{kj}$.
Therefore, pseudo-connection $\lambda_{ij}$ is decomposed as a sum of multi-step connections:
\begin{align}
	\lambda_{ij} \!=\!
	w_{ij}\!+\!\sum_{k\in V}\!w_{ik}\theta_{kj}\!+\!\sum_{k,l\in V}\!w_{ik}\theta_{kl}\theta_{lj}\!+\!\sum_{k,l,m\in V}\!w_{ik}\theta_{kl}\theta_{lm}\theta_{mj}\!+\!\cdots.
	\label{eq:lamb_dec}
\end{align}
Let $W, \Lambda, \Theta$ be matrices whose $(i, j)$ elements are $w_{ij}, \lambda_{ij}$, and $\theta_{ij}$, respectively.
Then, \Equref{eq:lamb_dec} is also represented in matrix form:
\begin{align}
	\Lambda = W(I + \Theta + \Theta^2 + \Theta^3+\cdots) = W(I - \Theta)^{-1}.
	\label{eq:est_W}
\end{align}

\subsection{Algorithm to estimate connections}
Now, connection $w_{ij}$ can be estimated from pseudo-connection $\lambda_{ij}$ by \Equref{eq:theta_w} and \Equref{eq:est_W}.
We need to obtain connection $w_{ij}$ that is consistent with two propagations: the short period one in \Equref{eq:theta_w} and the long period one in \Equref{eq:est_W}.
The problem is solved by alternating iterative procedures.

Pseudo-connection $\lambda_{ij}$ is estimated from spike trains by \Equref{eq:est_lambda}.
We can specify the pseudo-connection up to the unknown multiplicative factor $\rho$.
To estimate the relative strength of connections, we set $\rho=1$.
Let $W^{(\tau)}, \Lambda^{(\tau)}, \Theta^{(\tau)}$ be $W, \Lambda, \Theta$ in the $\tau$-th iteration.
For $\tau=0$, the elements of $\Theta$ are initialized by uniform random numbers in $[0, 1]$.
First, by fixing $\Lambda^{(\tau-1)}$ and $\Theta^{(\tau-1)}$, we estimate $W^{(\tau)}$ as
\begin{align}
  W^{(\tau)} = \Lambda^{(\tau-1)}(I - \Theta^{(\tau-1)}).
	\label{eq:alg_estW}
\end{align}
Second, we estimate $\Theta^{(\tau)}$ by fixing $W^{(\tau)}$ as
\begin{align}
  \bigl[\Theta^{(\tau)}\bigr]_{ij} = \Phi_1\bigl([W^{(\tau)}]_{ij} + \bar{C}_{ij}\bigr),
	\label{eq:alg_estT}
\end{align}
where $\bar{C}_{ij}$ is calculated by \Equref{eq:nonObs} from spike data.
The diagonal elements of $\Theta^{(\tau)}$ are replaced by $0$ to impose the assumption that a neuron does not have self-connection.
We also approximate self-pseudo-connection $\lambda_{ii}$ during a long period $[t_{\Delta}]$ as follows:
\begin{align}
	\lambda_{ii} \simeq \sum_{k\in V}\lambda_{ik}{\theta}_{ki}.
	\label{eq:arg_l}
\end{align}
The right-hand side of this equation represents the expectation of inputs via multiplicative neurons from neuron $k$ firing owing to the firing of neuron $i$.
In matrix form, \Equref{eq:arg_l} is represented by
\begin{align}
	\mathrm{diag}\bigl(\Lambda^{(\tau)}\bigr)
	&\leftarrow \mathrm{diag}\bigl(\Lambda^{(\tau-1)}\Theta^{(\tau)}\bigr),
	\label{eq:alg_diagL}
\end{align}
where $\mathrm{diag}(A)$ denotes the diagonal elements of matrix $A$.
These procedures are iterated for the pre-set number of iterations $M$.
Empirically, the result of estimating $W$ converges sufficiently when $M=10$.
The obtained $W^{(M)}$ provides the estimation of connection $W$.
These procedures are summarized in Algorithm~1.

The connections obtained by Algorithm~1 tend to yield neurons that have both excitatory and inhibitory influences on the target neurons.
This result is not consistent with the general neuroscience principle that a single neuron has only one type of connection.
To explicitly incorporate this requirement into the estimation procedure, we propose an improved algorithm in the next subsection.
\begin{algorithm}[h]
	\caption{Estimating connections $W$}
	\begin{algorithmic}[1]
		\State{\textbf{Inputs:}\;\;$\Lambda$, $\bar{C}$}
		\Function{estimateW}{$\Lambda$, $\bar{C}$}
		\State{\textbf{Initialization:}\;\;$\Theta^{(0)}\leftarrow[0, 1]^{N\times N}$，\;\;$\Lambda^{(0)}\leftarrow\Lambda$}
		\For{$\tau=1$ to $\tau=M$}
			\State{$W^{(\tau)} \leftarrow \Lambda^{(\tau-1)}(I - \Theta^{(\tau-1)})$}
			\Comment{\Equref{eq:alg_estW}}
			\For{$i=1$ to $i=N$}
				\For{$j=1$ to $j=N$}
					\State{$\bigl[\Theta^{(\tau)}\bigr]_{ij} \leftarrow \Phi_1\bigl([W^{(\tau)}]_{ij} + [\bar{C}]_{ij}\bigr)$}
					\Comment{\Equref{eq:alg_estT}}
				\EndFor
			\EndFor
			\State{$\mathrm{diag}(\Theta^{(\tau)})\leftarrow 0$}
			\State{$\mathrm{diag}\bigl(\Lambda^{(\tau)}\bigr) \leftarrow \mathrm{diag}\bigl(\Lambda^{(\tau-1)}\Theta^{(\tau)}\bigr)$}
			\Comment{\Equref{eq:alg_diagL}}
	\EndFor
	\EndFunction
	\State{\textbf{Output:}\;\;$W^{(M)}$}
	\end{algorithmic}
\end{algorithm}
\subsection{Algorithm to estimate connections on the basis of excitatory and inhibitory labels}
We propose an improved algorithm to estimate connections $W$ by pruning inconsistent connections on the basis of labels indicating whether a neuron is excitatory or inhibitory.
Suppose that we know the type of the observed neurons. Let $\bm{z}=(z_1, z_2, \cdots, z_N)\in\{1, 0\}^N$, where $z_i=1$ implies that neuron $i$ has only excitatory connections and $z_i=0$ implies that neuron $i$ has only inhibitory connections.
We refer to the binary vector $\bm{z}$ as an {\it excitatory-inhibitory label} or a {\it label}.
First, connections $W^{(\tau)}$ are estimated using \Equref{eq:alg_estW}.
Second, the inconsistent connections in $W^{(\tau)}$ are truncated on the basis of label $\bm{z}$.
This yields a new estimator $W[\bm{z}]^{(\tau)}$, which is consistent with the neuroscience principle that a single neuron has only one type of connection.
The procedure is implemented as follows. If neuron $j$ is excitatory~($z_j=1$), the estimated inhibitory connections from neuron $j$ are truncated as
\begin{align}
  \Bigl[W[\bm{z}]^{(\tau)}\Bigr]_{ij} \leftarrow
  \max{\Bigl([W^{(\tau)}]_{ij}, 0\Bigr)}.
	\label{eq:proj_exc}
\end{align}
If neuron $j$ is inhibitory~($z_j=0$), the estimated excitatory connections from neuron $j$ are truncated as
\begin{align}
  \Bigl[W[\bm{z}]^{(\tau)}\Bigr]_{ij} \leftarrow
  \min{\Bigl([W^{(\tau)}]_{ij}, 0\Bigr)}.
  \label{eq:proj_inh}
\end{align}
Then, we update the parameters $\Theta$ by fixing $W[\bm{z}]^{(\tau)}$ and replace the diagonal elements of $\Theta^{(\tau)}$ and $\Lambda^{(\tau-1)}$ in the same way as in Algorithm 1.
These procedures are iterated for the pre-set number of iteration $M$.
We obtain $W[\bm{z}]^{(M)}$ as the estimation of connections $W$.
These procedures are summarized in Algorithm~2.

However, in general, we do not know whether a neuron is excitatory or inhibitory in advance.
Therefore, we develop an information-theoretic framework for estimating neuron types.
\begin{algorithm}[H]
	\caption{Estimating connections $W$ given excitatory-inhibitory labels}
	\begin{algorithmic}[1]
		\State{\textbf{Inputs:}\;\;$\Lambda$, $\bar{C}$, $\bm{z}$}
		\Function{estimateW\_label}{$\Lambda$, $\bar{C}$, $\bm{z}$}
		\State{\textbf{Initialization:}\;\;$\Theta^{(0)}\leftarrow[0, 1]^{N\times N}$，\;\;$\Lambda^{(0)}\leftarrow\Lambda$}
		\For{$\tau=1$ to $\tau=M$}
			\State{$W^{(\tau)} \leftarrow \Lambda^{(\tau-1)}(I - \Theta^{(\tau-1)})$}
			\Comment{\Equref{eq:alg_estW}}
			\For{$i=1$ to $i=N$}
				\For{$j=1$ to $j=N$}
					\If{$z_j=1$}
          \State{$\Bigl[W[\bm{z}]^{(\tau)}\Bigr]_{ij}\leftarrow\max{\Bigl([W^{(\tau)}]_{ij},0\Bigr)}$}
          \Comment{\Equref{eq:proj_exc}}
					\ElsIf{$z_j=0$}
          \State{$\Bigl[W[\bm{z}]^{(\tau)}\Bigr]_{ij}\leftarrow\min{\Bigl([W^{(\tau)}]_{ij},0\Bigr)}$}
          \Comment{\Equref{eq:proj_inh}}
					\EndIf
				\EndFor
			\EndFor
			\For{$i=1$ to $i=N$}
				\For{$j=1$ to $j=N$}
					\State{$\bigl[\Theta^{(\tau)}\bigr]_{ij} \leftarrow \Phi_1\bigl([W^{(\tau)}]_{ij} + \bar{C}_{ij}\bigr)$}
					\Comment{\Equref{eq:alg_estT}}
				\EndFor
			\EndFor
			\Comment{\Equref{eq:alg_estT}}
			\State{$\mathrm{diag}(\Theta^{(\tau)})\leftarrow 0$}
			\State{$\mathrm{diag}\bigl(\Lambda^{(\tau)}\bigr) \leftarrow \mathrm{diag}\bigl(\Lambda^{(\tau-1)}\Theta^{(\tau)}\bigr)$}
			\Comment{\Equref{eq:alg_diagL}}
	\EndFor
	\EndFunction
	\State{\textbf{Output:}\;\;$W[\bm{z}]^{(M)}$}
	\end{algorithmic}
\end{algorithm}

\section{Estimation of excitatory probability}
In this section, we describe an information-theoretic framework that classifies a neuron as an excitatory or inhibitory neuron.
The framework uses the $em$ algorithm to obtain the optimal excitatory-inhibitory label.

\subsection{Model}
We assume that a binary random variable $z_i\in\{1, 0\}$ follows a Bernoulli distribution with parameter $\alpha_i$ and that the variables $z_i\;(i=1, \cdots, N)$ are mutually independent.
Therefore, a random variable vector $\bm{z}\in\{1, 0\}^{N}$ follows the distribution of the form~:
\begin{align}
	p'(\bm{z}; \bm{\alpha}) = \prod_{i=1}^{N}\alpha_i^{z_i}(1 - \alpha_i)^{(1 - z_i)},
	\label{eq:factBer_a}
\end{align}
where $\bm{z} = (z_1, \cdots, z_N)^{\mathrm T}$ and $\bm{\alpha} = (\alpha_1, \cdots, \alpha_N)^{\mathrm T}$.
In this study, we estimate the parameter vector $\bm{\alpha}$ by the $em$ algorithm~\citep{amari1995information}, which is widely used to identify a model including latent variables.
If the estimated parameter $\alpha_i$ is larger than the pre-set threshold, neuron $i$ is classified as excitatory and vice versa.
How to set the threshold will be discussed later.
Hereafter, we refer to the parameter $\alpha_{i}$ as the {\it excitatory probability}.

\subsection{$em$ algorithm}
Consider a probability distribution space $\mathcal{S}$.
The KL divergence $D_\mathrm{KL}$ between two distributions $p(x)$ and $q(x)$ is defined by
\begin{align}
	D_\mathrm{KL}(p, q) = \sum_{x} p(x)\log{\frac{p(x)}{q(x)}}.
\end{align}
In this space, two geodesics can be defined, namely $e$-geodesic and $m$-geodesic, which are natural in terms of the KL divergence.
The $m$-geodesic in $\mathcal{S}$ is the set of internal divisions between two distributions.
The $m$-geodesic connecting the distributions $p(x)$ and $q(x)$ in $\mathcal{S}$ is defined by
\begin{align}
	r(x, t) = (1-t) \cdot p(x) + t \cdot q(x), \qquad 0\leq t \leq1,
\end{align}
where $t$ is a parameter.
On the other hand, the $e$-geodesic in $\mathcal{S}$ is the set of internally dividing points between two distributions in logarithmic form.
The $e$-geodesic connecting the distributions $p(x)$ and $q(x)$ in $\mathcal{S}$ is defined by
\begin{align}
	\log{r(x, t)} = (1-t) \cdot \log{p(x)} + t \cdot \log{q(x)} - \psi(t), \qquad 0\leq t \leq1,
\end{align}
where $\psi(t)$ is a normalization factor that ensures that points $r(x, \bm{t})$ are probability distributions.
Similar to extending a line to a surface, an $m$-flat subspace $\mathcal{M}_m$ consisting of $K$ distributions $\{p_k(x)\}_{k=1, \cdots, K}$ is defined by
\begin{align}
	\mathcal{M}_m = \Biggl\{r(x, \bm{t}) = \sum_{k=1}^K t_k p_k(x), t_k > 0, \sum_{k=1}^K t_k =1\Biggr\},
\end{align}
and an $e$-flat subspace $\mathcal{M}_e$ is defined by
\begin{align}
  \mathcal{M}_e = \Biggl\{r(x, \bm{t}) = \exp\Bigl(\sum_{k=1}^K t_k \log{p_k(x)} - \psi(\bm{t}), t_k > 0, \sum_{k=1}^K t_k =1\Bigr)\Biggr\}.
\end{align}
Then, we define the projection of a point $p$ in $\mathcal{M}_m$ to $\mathcal{M}_e$ by
\begin{align}
	r_m = \argmin_{r\in\mathcal{M}_e} D_\mathrm{KL}(p, r),
\end{align}
which is called the $m$-projection.
We also define the projection of a point $q$ in $\mathcal{M}_e$ to $\mathcal{M}_m$ by
\begin{align}
	r_e = \argmin_{r\in\mathcal{M}_m} D_\mathrm{KL}(r, q),
\end{align}
which is called the $e$-projection.
The $em$ algorithm consists of two projection steps: the $e$-step and the $m$-step.
We note that our model manifold $\mathcal{M}$ has the $e$-flat structure and the observed data manifold $\mathcal{D}$ has the $m$-flat structure.
By iterating $e$- and $m$-steps alternately, the $em$ algorithm finds a point in $\mathcal{M}\subset\mathcal{M}_e$ that is the closest to $\mathcal{D}\subset\mathcal{M}_m$.
The projection steps of the $em$ algorithm are shown in \Figref{fig:em_step} (left).

\subsection{Estimation of excitatory probability using the $em$ algorithm}
\begin{figure}[h]
 	\centering
 	\includegraphics[bb=0 0 950 381, trim=0 0 0 0, width=\textwidth]{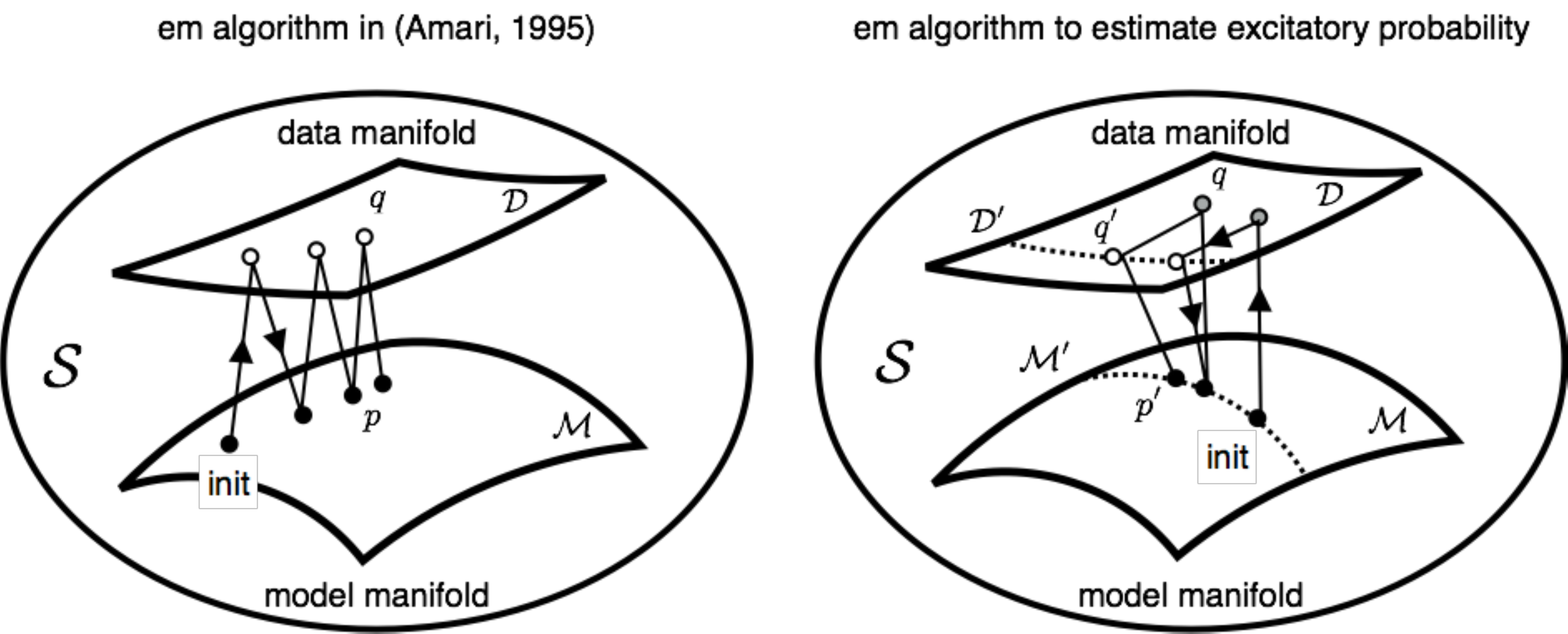}
   \caption{The $em$ algorithm in \citet{amari1995information} is on the left and the $em$ algorithm to estimate excitatory probability is on the right.
   In the $em$ algorithm on the left, the projections are iterated between the dual spaces.
   In our $em$ algorithm on the right, the projections are iterated between subspaces restricted in the factorial model on the dual spaces~(dashed line).}
 	\label{fig:em_step}
 \end{figure}

In this study, the pseudo-connections $\Lambda$ are regarded as observed random variables and the labels $\bm{z}$ are regarded as latent variables.
Let the set of the empirical marginal distribution of $\Lambda$ be a data manifold $\mathcal{D}$, which is an $m$-flat subspace.
Any point on $\mathcal{D}$ represents the following distribution:
\begin{align}
  q(\Lambda, \bm{z}; \bm{\beta})
  = q(\Lambda)q(\bm{z}\mid\Lambda; \bm{\beta}),
\end{align}
where $q(\Lambda)$ is the empirical distribution of $\Lambda$ and $\bm{\beta}$ is a parameter vector with $\bm{z}$ similar to the parameter $\bm{\alpha}$.
On the other hand, a model manifold $\mathcal{M}$ is an $e$-flat subspace; hence, any point on $\mathcal{M}$ represents the following distribution:
\begin{align}
  p(\Lambda, \bm{z}; \bm{\alpha})
  = p(\bm{z}; \bm{\alpha})p(\Lambda\mid\bm{z}; W).
\end{align}
We assume that the distribution $p(\bm{z}; \bm{\alpha})$ is a factorization model of a Bernoulli distribution as \Equref{eq:factBer_a}.
The factorial distribution $p'(\Lambda, \bm{z}; \bm{\alpha})\Bigl(=p'(\bm{z}; \bm{\alpha})p(\Lambda\mid\bm{z}; W)\Bigr)$ consists of the $e$-flat subspace $\mathcal{M}' \subset\mathcal{M}$.
We also assume that the distribution of label $\bm{z}$ given observation $\Lambda$ is a factorial model:
\begin{align}
	q'(\bm{z}\mid\Lambda; \bm{\beta})
	= \prod_{i=1}^{N}\beta_i^{z_i}(1 - \beta_i)^{(1 - z_i)}.
\end{align}
The distribution $q'(\Lambda, \bm{z}; \bm{\beta})\Bigl(=q(\Lambda)q'(\bm{z}\mid\Lambda; \bm{\beta})\Bigr)$ consists of the $e$-flat subspace $\mathcal{D}'\subset\mathcal{D}$.
Therefore, we estimate the parameter $\bm{\alpha}$ by finding a distribution $p'\in\mathcal{M}'$ that is the closest to $\mathcal{D}'$.
We update two parameters, namely $\bm{\alpha}$ and $\bm{\beta}$, as shown in \Figref{fig:em_step} (right).
First, the parameter $\bm{\beta}$ is updated by $e$-projection from $\mathcal{M}'$ to $\mathcal{D}$ and $e$-projection from $\mathcal{D}$ to $\mathcal{D}'$.
Then, the parameter $\bm{\alpha}$ is updated by $m$-projection from $\mathcal{D}'$ to $\mathcal{M}'$.
These procedures are iterated alternately until a condition with the parameter $\bm{\alpha}$ is satisfied.
The projection from $\mathcal{M}'$ to $\mathcal{D}'$ cannot be obtained analytically; therefore, we modified the procedures from the original $em$ algorithm in \citet{amari1995information}.
The iterative approach is described below.

\subsubsection{Projection from $\mathcal{M}'$ to $\mathcal{D}'$}
The projection from the model subspace $\mathcal{M}'$ to the data subspace $\mathcal{D}'$ consists of two inner steps.

\noindent
{\bf Step~1.} A point $p'\in\mathcal{M}'$ is projected to a point $q\in\mathcal{D}$ by $e$-projection as
\begin{align}
  \bm{\beta}^{(\tau+1)} = \argmin_{\bm{\beta}}D_{\mathrm KL}[q(\Lambda, \bm{z}; \bm{\beta}), p'(\Lambda, \bm{z}; \bm{\alpha}^{(\tau)})].
\end{align}
Practically, the distribution $q(\bm{z}\mid \Lambda; \bm{\beta})$ is substituted by the distribution $p'(\bm{z}\mid \Lambda; \bm{\alpha}^{(\tau)})$ on the basis of the property of $e$-projection given in \citet{amari1995information}:
\begin{align}
  q(\bm{z}\mid \Lambda; \bm{\beta}) \leftarrow p'(\bm{z}\mid \Lambda; \bm{\alpha}^{(\tau)}).
	\label{eq:step1}
\end{align}
By Bayes' theorem, the right-hand side of \Equref{eq:step1} is calculated by
\begin{align}
	p'(\bm{z}\mid \Lambda; \bm{\alpha}^{(\tau)})
	= \frac{p(\Lambda\mid \bm{z}; W)p'(\bm{z}; \bm{\alpha}^{(\tau)})}{\sum_{\bm{z}}p(\Lambda\mid \bm{z}; W)p'(\bm{z}; \bm{\alpha}^{(\tau)})},
	\label{eq:step1_bayes}
\end{align}
where $p(\Lambda\mid \bm{z}; W)$ is the likelihood of observation $\Lambda$ given the truncated estimator $\hat{\Lambda}[z] \Bigl(=W[\bm{z}](I - \Theta)^{-1}\Bigr)$, and we assume that the error in each pseudo-connection follows a Gaussian distribution with mean $0$ and variance $\sigma_e^2$.
Then, the distribution $p(\Lambda\mid \bm{z}; W)$ is calculated by
\begin{align}
	p(\Lambda\mid \bm{z}; W)
  &= \prod_{i,j}\Biggr[
		\frac{1}{\sqrt{2\pi\sigma_e^2}}
		\exp\Bigl\{
		-\frac{(\lambda_{ij}-\hat{\lambda}_{ij})^2}{2\sigma_e^2}
		\Bigr\}\Biggl]
  \nonumber\\
  &= \prod_{i,j,\lambda_{ij}>0}\Biggr[
    \frac{1}{\sqrt{2\pi\sigma_e^2}}
		\exp\Bigl\{
		-\frac{(\lambda_{ij}-\hat{\lambda}_{ij})^2}{2\sigma_e^2}
		\Bigr\}\Biggl]
  \nonumber\\
  &\qquad\qquad\times
  \prod_{i,j,\lambda_{ij}<0}\Biggr[
		\frac{1}{\sqrt{2\pi\sigma_e^2}}
		\exp\Bigl\{
		-\frac{(\lambda_{ij}-\hat{\lambda}_{ij})^2}{2\sigma_e^2}
		\Bigr\}\Biggl]
  \nonumber\\
  &\simeq \prod_{i,j,\lambda_{ij}>0}\Biggr[
    \frac{1}{\sqrt{2\pi\sigma_e^2}}
		\exp\Bigl\{
		-\frac{(\lambda_{ij}-\hat{\lambda}_{ij})^2}{2\sigma_e^2}
		\Bigr\}\Biggl].
	\label{eq:likelihood}
\end{align}
In the above expansion, we assume that $\sigma_e\gg0$ if $\lambda_{ij}<0$.

\noindent
{\bf Step~2.} A point $q\in\mathcal{D}$ is projected to a point $q'\in\mathcal{D}'$ by $e$-projection as
\begin{align}
	\bm{\beta}^{(\tau+1)} = \argmin_{\bm \beta} D_{\mathrm{KL}}[q'(\Lambda, \bm{z}; \bm{\alpha}^{(\tau)}), q(\Lambda, \bm{z}; \bm{\beta})].
\end{align}
Hence, the parameter $\bm{\beta}$ is updated by
\begin{align}
	\beta_i^{(\tau+1)} &\leftarrow \cfrac{1}{1+\exp{\bigl\{-A^{(\tau)}\bigr\}}},
	\\
	A^{(\tau)} &= \sum_{\bm{z}_{-i}} q'(\bm{z}_{-i}\mid\Lambda;\beta) \log{\cfrac{p'(\bm{z}_{-i}\mid\Lambda, z_i=1; \bm{\alpha}^{(\tau)})}{p'(\bm{z}_{-i}\mid\Lambda, z_i=0; \bm{\alpha}^{(\tau)})}},
	\label{eq:step2}
\end{align}
where $\bm{z}_{-i}=(z_1, \cdots, z_{i-1}, z_{i+1}, \cdots, z_{N})$ and $q'(\bm{z}_{-i}\mid\Lambda;\beta)$ is given by $q'(\bm{z}_{-i}\mid\Lambda;\beta) = \prod_{j\neq i} \beta_j^{z_j}(1-\beta_j)^{1-z_j}$.
In this step, $q'\in\mathcal{D}'$, the closest point to $q\in\mathcal{D}$, needs to be obtained by projection.
Both $e$-projection and $m$-projection can be used in this step because the KL divergence between $q'\in\mathcal{D}'$ and $q\in\mathcal{D}$ is small.
In this study, $e$-projection is adopted.
The subspace $\mathcal{D}'$ is $e$-flat; hence, the $e$-projection to $\mathcal{D}'$ is not uniquely determined in general.
However, it is experimentally confirmed that there is no significant difference between the estimations of the excitatory probability $\bm{\alpha}$ by $e$-projection and $m$-projection in this step.

\subsubsection{Projection from $\mathcal{D}'$ to $\mathcal{M}'$}
The $m$-projection of the distribution $q'\in\mathcal{D}'$ to $\mathcal{M}'$ gives the distribution $p'$, i.e.,
\begin{align}
	\bm{\alpha}^{(\tau+1)} = \argmin_{\bm \alpha} D_{\mathrm{KL}}[q'(\Lambda, \bm{z}; \bm{\beta}^{(\tau)}), p(\Lambda, \bm{z}; \bm{\alpha})].
\end{align}
Therefore, parameter $\bm{\alpha}$ is updated by
\begin{align}
	\alpha_i^{(\tau+1)} \leftarrow  \sum_{\bm z}z_iq'(z_i \mid \Lambda; \bm{\beta}^{(\tau)}).
	\label{eq:step3}
\end{align}
The derivation of the $em$ algorithm to estimate parameter $\bm{\alpha}$ is given in Appendix~B.

These two steps are iterated until parameter $\bm{\alpha}$ converges.
The proposed algorithm is summarized in Algorithm~3.
If the estimated value of $\alpha_i$ is equal to $1$, neuron $i$ is classified as excitatory; otherwise, neuron $i$ is classified as inhibitory.
\begin{algorithm}[h]
 	\caption{$em$ algorithm to estimate excitatory probability}
 	\begin{algorithmic}[1]
    \State{\textbf{Inputs:}~$\Lambda$, $\hat{C}$}
 		\State{\textbf{Initialization:}~${\alpha}_i^{(1)} \leftarrow \frac{1}{2},\;\; i=1, 2, \cdots, N$}
 		\While{until convergence}
 		\ForAll{$\bm{z}\in\{0, 1\}^N$}
 			\State{$W(\bm{z})
 				\leftarrow \Call{estimateW\_label}{\Lambda, \bar{C}, \bm{z}}$}
      \State{$\hat{\Lambda}=W[\bm{z}](I-\Theta)^{-1}$}
 			\State{$p(\Lambda\mid \bm{z})=\prod_{i,j}\Bigr[\frac{1}{\sqrt{2\pi\sigma_e^2}}\exp\Bigl\{-\frac{(\lambda_{ij}-\hat{\lambda}_{ij})^2}{2\sigma_e^2}\Bigr\}\Bigl]$}
 			\Comment{\Equref{eq:likelihood}}
 		\EndFor
 		\State{$q(\bm{z}\mid\Lambda; \bm{\alpha}^{(\tau)}) \leftarrow \frac{p(\Lambda\mid\bm{z})p(\bm{z}; \bm{\alpha}^{(\tau)})}{\sum_{\bm{z}}p(\Lambda\mid\bm{z})p(\bm{z}; \bm{\alpha}^{(\tau)})}$}
 		\Comment{\Equref{eq:step1}}
    \State{$A^{(\tau)} = \sum_{\bm{z}_{-i}} q'(\bm{z}_{-i}\mid\Lambda;\beta) \log{\frac{p(\bm{z}_{-i}\mid\Lambda, z_i=1; \bm{\alpha}^{(\tau)})}{p(\bm{z}_{-i}\mid\Lambda, z_i=0; \bm{\alpha}^{(\tau)})}}$}
    \Comment{\Equref{eq:step2}}
    \State{$\beta_i^{(\tau+1)} \leftarrow \frac{1}{1+\exp{\bigl\{-A^{(\tau)}\bigr\}}}$}
 		\State{$\alpha_i^{(\tau+1)} \leftarrow  \sum_{\bm z}z_iq'(z_i\mid\Lambda; \bm{\beta}^{(\tau+1)})$}
 		\Comment{\Equref{eq:step3}}
 	\EndWhile
 	\State{\textbf{Output:}~$\bm{\alpha}$}
 	\end{algorithmic}
\end{algorithm}
\subsubsection{Approximated algorithm by Gibbs sampling}
Algorithm 3 is computationally expensive because of the combinatorial explosion.
The variables $\bm{z}\in\{1, 0\}^N$ have $2^N$ combinations.
The computational costs of \Equref{eq:step1}, \Equref{eq:step2}, and \Equref{eq:step3} rapidly increase with the number of neurons $N$.
To overcome this difficulty, we approximate \Equref{eq:step2} by Gibbs sampling.
Gibbs sampling is a Markov chain Monte Carlo (MCMC) sampling technique that is used for computing the statistics of a target distribution approximately when direct sampling from the distribution is difficult.
Here, the target distribution is $q'$ in \Equref{eq:step2}.
Suppose that we generate $K$ samples from distribution $q'$.
The $k$-th sample $z_i^{(k)}$ of $z_i$ is sampled by
\begin{align}
	z_i^{(k+1)}
	\sim q'(z_i \mid z_1^{(k)}, \cdots z_{i-1}^{(k)}, z_{i+1}^{(k)}, \cdots, z_N^{(k)}, \Lambda; \bm{\alpha}).
	\label{eq:gibbs_samp}
\end{align}
From \Equref{eq:step1}, \Equref{eq:gibbs_samp} is given by
\begin{align}
	z_i^{(k+1)} \sim \frac{p_1}{p_1 + p_0},
\end{align}
where $p_1$ and $p_0$ are calculated by
\begin{align}
	p_1 &= p(\Lambda\!\mid\! z_i\!=\!1, \bm{z}_{-i}^{(k)}; W)p(z_i\!=\!1, \bm{z}_{-i}^{(k)}; \bm{\alpha}^{(\tau)}),
	\\
	p_0 &= p(\Lambda\!\mid\! z_i\!=\!0, \bm{z}_{-i}^{(k)}; W)p(z_i\!=\!0, \bm{z}_{-i}^{(k)}; \bm{\alpha}^{(\tau)}),
\end{align}
where $\bm{z}_{-i}^{(k)}=\{z_1^{(k)}, \cdots z_{i-1}^{(k)}, z_{i+1}^{(k)}, \cdots, z_N^{(k)}\}$.
By using sampled $\{\bm{z}^{(k)}\}_{k=1, \cdots, K}$, we approximate \Equref{eq:step1} and \Equref{eq:step2} by
\begin{align}
  \bm{\beta}_i^{(\tau+1)} \leftarrow \frac{1}{K}\sum_{k=1}^{K} q'(\bm{z}_{-i}^{(k)}\mid\Lambda;\beta) \log{\cfrac{p(\bm{z}_{-i}^{(k)}\mid\Lambda, z_i=1; \bm{\alpha}^{(\tau)})}{p(\bm{z}_{-i}^{(k)}\mid\Lambda, z_i=0; \bm{\alpha}^{(\tau)})}}.
	\label{eq:gibbs_samp2}
\end{align}
\Equref{eq:step3} is calculated by sampling $\bm{z}$ randomly.
We let the estimator of $\bm{\alpha}$ be the mean of $M_{\bm{\alpha}}$ estimates of $\bm{\alpha}$ by using the $em$ algorithm:
\begin{align}
  \bar{\alpha}_i = \frac{1}{M_{\bm{\alpha}}}
  \sum_{m=1}^{M_{\bm{\alpha}}} \alpha_{i}^{m},
\end{align}
where $\alpha_{i}^{m}$ is a parameter of the $m$-th label estimation.
If the value of $\bar{\alpha}_i$ is equal to $1$, neuron $i$ is classified as excitatory; otherwise, neuron $i$ is classified as inhibitory.

\section{Numerical simulations}
In this section, the proposed methods are verified using synthesized data.

\subsection{Spiking model}
We generated synthetic data using a simple spiking model proposed by \citet{Izhikevich2003simple}.
This model is known to be able to simulate various types of cortical neuron behaviors.
Following \citet{Izhikevich2003simple}, we used regular spiking cells to model excitatory neurons and fast spiking cells to model inhibitory neurons.
The four parameters that specify a neuron's behavior are summarized in Table \ref{tb:izh_param}.
The variables $r_e$ and $r_i$ are random variables from a uniform distribution with the interval $[0, 1]$.
\begin{table}[h]
	\centering
  \caption{Parameters that specify a neuron's behavior in \citet{Izhikevich2003simple}}
	\begin{tabular}{r|c|c|c|c}
		\hline
		& $a$ & $b$ & $c$ & $d$ \\ \hline
		excitatory & $0.02$ & $0.2$ & $-65 + 15 r_e^2$ & $ 8 - 6 r_e^2$ \\ \hline
		inhibitory & $0.02 + 0.08r_i$ & $0.25 - 0.05r_i$ & $-65$ & $2$\\ \hline
	\end{tabular}
	\label{tb:izh_param}
\end{table}

\subsection{Synthetic data}
\subsubsection{Network of 100 neurons}
To investigate the performance of the proposed method in a simple setting, we constructed a network of 100 neurons~($N_A=100$).
On the basis of neuroscience evidence, $80$ neurons were excitatory and $20$ neurons were inhibitory, and the number of connections from one neuron to others was set to at most $10$ (corresponding to a $10\%$ connection ratio).
Their excitatory and inhibitory connect weights follow uniform distributions $[0, 10]$ and $[-10, 0]$, respectively.
To investigate the performance for partial observations, we randomly sampled $N=33$ neurons and constructed a connection matrix $W^*\in\mathbb{R}^{N\times N}$ from $N_A=100$ neurons.
The number of unobserved neurons was set to be twice as high as that of observed ones.
A network activity was simulated for 1 h with a temporal resolution of 1 ms.

\subsubsection{Network of 10,000 neurons}
\begin{figure}[h]
  \begin{minipage}{0.5\hsize}
    \centering
    \includegraphics[bb=0 0 504 504, width=\linewidth, trim=50 50 50 50]{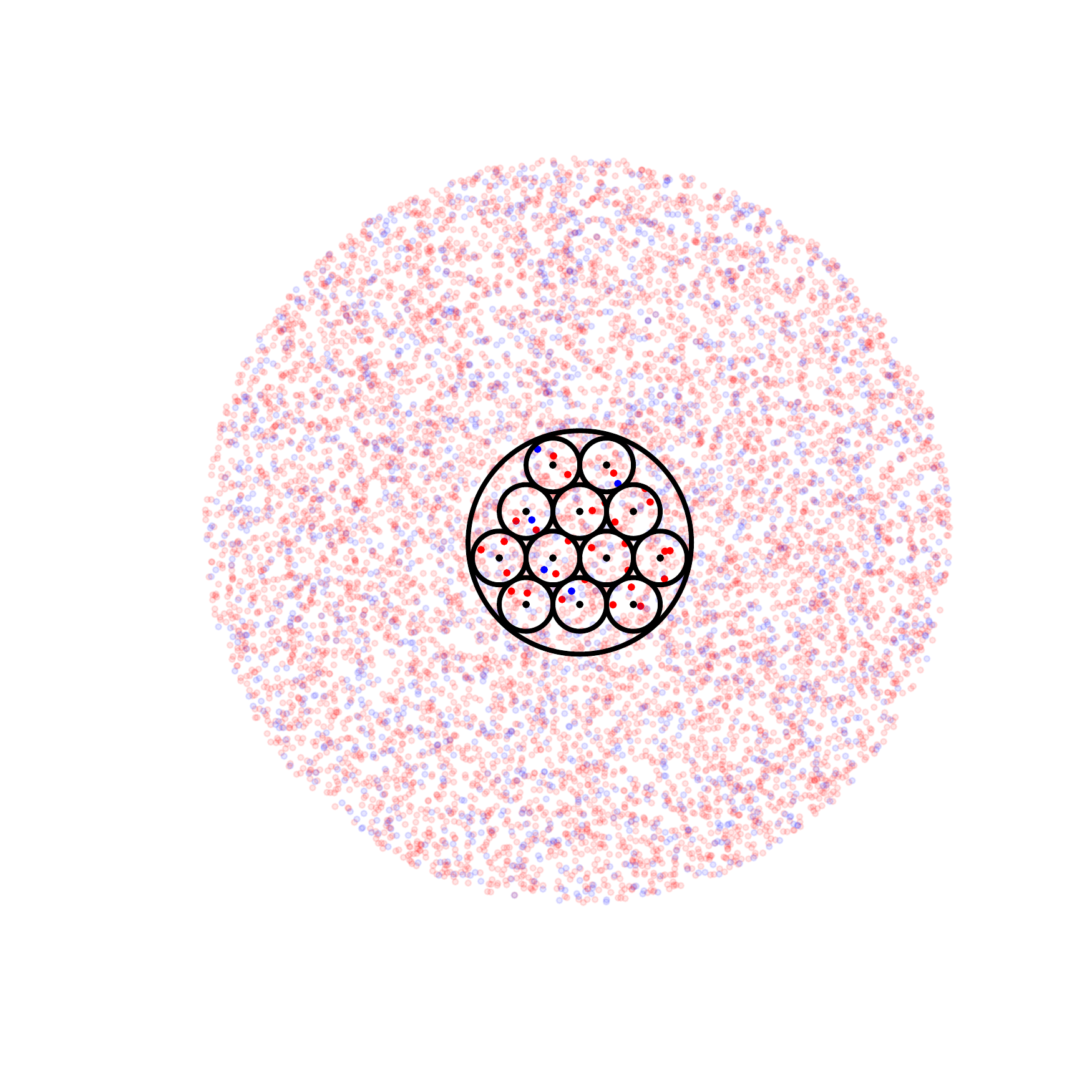}
  \end{minipage}
  \begin{minipage}{0.5\hsize}
    \centering
    \includegraphics[bb=0 0 504 504, width=\linewidth]{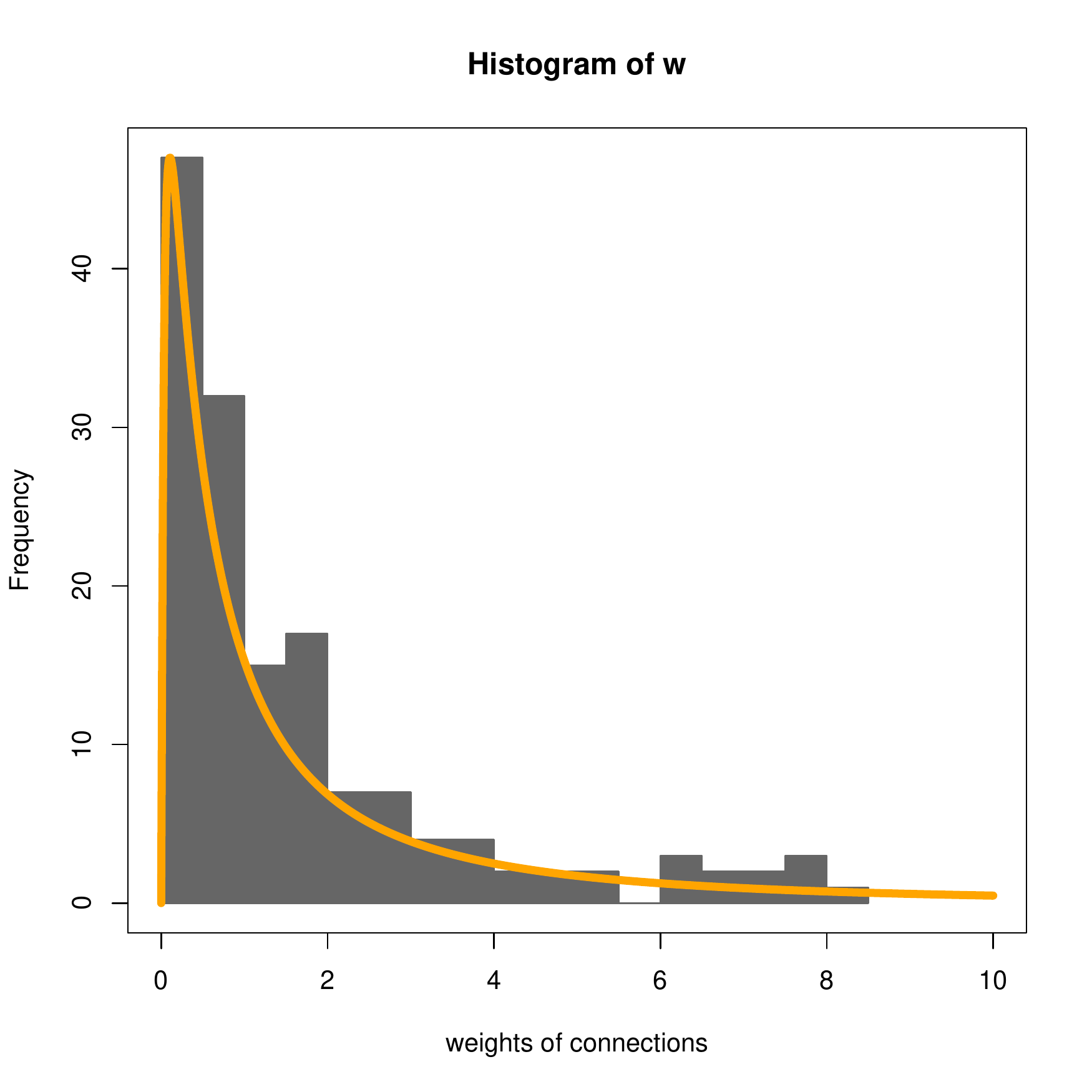}
  \end{minipage}
  \caption{Schematic illustration of $10,000$-neuron simulation and distribution of the connection weights.
  Left: Positions of 10,000 neurons and sampled neurons and 12 tetrodes.
  The light red and blue dots represent excitatory and inhibitory neurons, respectively.
  The 12 small black circles represent the recording range of the tetrodes.
  The dark red and blue dots represent the sampled neurons.
  Right: Histogram of connection weights. The orange line represents the log-normal distribution (mean $0$ and variance $1.5^2$) from which the connection weights were sampled.}
  \label{fig:synthesized_data}
\end{figure}

To evaluate how the proposed method works under a realistic condition, we also constructed a network of $10,000$ neurons ($N_A=10,000$) in which $8,000$ neurons are excitatory and $2,000$ neurons are inhibitory.
The neurons are scattered on a 2-D plane and the connection weights follow a log-normal distribution \citep{song2005highly}, as shown in \Figref{fig:synthesized_data}.
The number of connections of each neuron is set to 150.
The details of how the network was generated are given in Appendix C.
To investigate the performance for partial observation, we randomly sampled $N=33$ neurons out of $N_A=10,000$ neurons and constructed a connection matrix $W^*\in\mathbb{R}^{N\times N}$.
In multi-electrode recording, such as hyper-drive recording \citep{gothard1996dynamics}, 12 tetrodes (TTs) are arranged as shown in \Figref{fig:synthesized_data}.
The small inner circles represent the recording range of each TT and the large outer circle represents the entire hyper-drive recording range, which is approximately 1 mm in diameter.
A network activity was simulated for 1 h with a temporal resolution of 1 ms.

\subsection{Comparative methods}
We compared the proposed method with two commonly used methods, namely the cross-correlation function and binary logistic regression.

\subsubsection{Cross-correlation function}
The cross-correlation function is often used to estimate neural interactions~\citep{wilson1994reactivation, bartho2004characterization}.
A cross-correlation function between neurons $i$ and $j$ for a time lag $s\in\mathbb{N}$, $\mathrm{CC}_{ij}(s)$ is defined as
\begin{align}
  \mathrm{CC}_{ij}(s) =
  \sum_{t=1}^{T} X_i(t)X_j(t+s).
\end{align}
Connection weight $w_{ij}$ from neuron $j$ to neuron $i$ is estimated by
\begin{align}
  w_{ij} = \max_{s}{\cfrac{\mathrm{CC}_{ij}(s)}{\sum_{s=1}^{\Delta}\left|\mathrm{CC}_{ij}(s)\right|}},
\end{align}
where the value $\Delta$ is the maximum time lag, which is the same as the time window $\Delta$ used in the proposed method.

\subsubsection{Binary logistic regression with $L_1$ regularization}
Binary logistic regression is a standard statistical method to estimate the probability that a response variable is 1 and to assess an unobserved input.
To estimate the connections to neuron $i$, let a variable set
\begin{align}
  \bm{X}_{-i}[t_\Delta]=\{X_1[t_\Delta], \cdots, X_{i-1}[t_\Delta], X_{i+1}[t_\Delta], \cdots, X_N[t_\Delta]\}
\end{align}
consist of explanatory variables and let a variable $X_i(t)$ be a response variable.
The estimated probability $p_i$ is given by
\begin{align}
  p_i(t) = \mathrm{Pr}(X_i(t)=1\mid\bm{X}_{-i}[t_{\Delta}])
  = \frac{1}{1+\exp{\Bigl(-b_i-\bm{w}_i^{\mathrm{T}}\bm{X}_{-i}[t_{\Delta}]\Bigr)}},
\end{align}
where $b_i$ is a parameter representing an unobserved input and $\bm{w}_i\in\mathbb{R}^{N-1}$ is a coefficient.
The $j$-th element $w_{ij}$ of $\bm{w}_i$ represents the connection weight from neuron $j$ to neuron $i$.
The parameters $b_i$ and $\bm{w}_i$ are estimated by maximizing the log likelihood function with $L_1$ regularization:
\begin{align}
  b_i, \bm{w}_i = \argmin_{b_i, \bm{w}_i} \Bigl\{-\ell(b_i, \bm{w}_i) + \lambda\sum_{j=1}^{N-1} |w_{ij}| \Bigr\},
\end{align}
where $\lambda$ is a regularization parameter, which is determined by 10-fold cross-validation in this study.
The log-likelihood function $\ell(b_i, \bm{w}_i)$ is given by
\begin{align}
  \ell(b_i, \bm{w}_i) = \sum_{t=\Delta}^{T}\Bigl\{
    \log{p_i(t)} + \bigl(1-X_i(t)\bigr)\Bigl(-b_i-\bm{w}_i^{\mathrm T}\bm{X}_{-i}[t_\Delta]\Bigr)
  \Bigr\}.
\end{align}

\subsection{Evaluation of estimated $W$}
The performance of the proposed method was assessed by comparing the estimated graph $W$ and the true graph $W^*$.
We used two evaluation indices, namely {\it sensitivity} and {\it Kendall rank correlation coefficient}.
\subsubsection{Sensitivity}
The true positive (TP) is defined by the number of connections that satisfy both $w_{ij}\neq0$ and $w_{ij}^*\neq0$, and the false negative (FN) is defined by the number of connections that satisfy $w_{ij}=0$ and $w_{ij}^*\neq0$. The sensitivity is calculated by
\begin{align}
	\mathrm{Senstivity}
	= \frac{{\rm TP}}{{\rm TP} + {\rm FN}}\in[0, 1].
\end{align}
The sensitivity gives the rate of correctly estimated connections.

\subsubsection{Kendall rank correlation coefficient}
If $W^*$ and $W$ are vectorized as ${\bm w}^*\in\mathbb{R}^{N^2}$ and ${\bm w}\in\mathbb{R}^{N^2}$, $w^*_i$ and $w_i$ are the $i$-th elements of ${\bm w}^*$ and $\bm{w}$, respectively.
The Kendall rank correlation coefficient $\tau_{\mathrm{K}}$ is defined by
\begin{align}
	\tau_{\mathrm{K}}
	&= \frac{\sum_{i, j}^N Q_{ij}}{{}_NP_2},
	\\
	Q_{ij}
	&= \mathrm{sgn}(w^*_i - w_i)\cdot\mathrm{sgn}(w^*_j - w_j).
\end{align}
$\tau_{\mathrm{K}}=1$ is obtained when the method estimates the rank of connections exactly.
Note that the Kendall rank correlation coefficient requires $W$ and $W^*$ to have the same number of connections.
Therefore, we calculate the Kendall rank correlation coefficient by choosing the connections included in both $W$ and in $W^*$.

\subsection{Results}
\subsubsection{Network of 100 neurons}
\begin{figure}[H]
	\centering
	\includegraphics[bb=0 0 921 345, width=\linewidth]{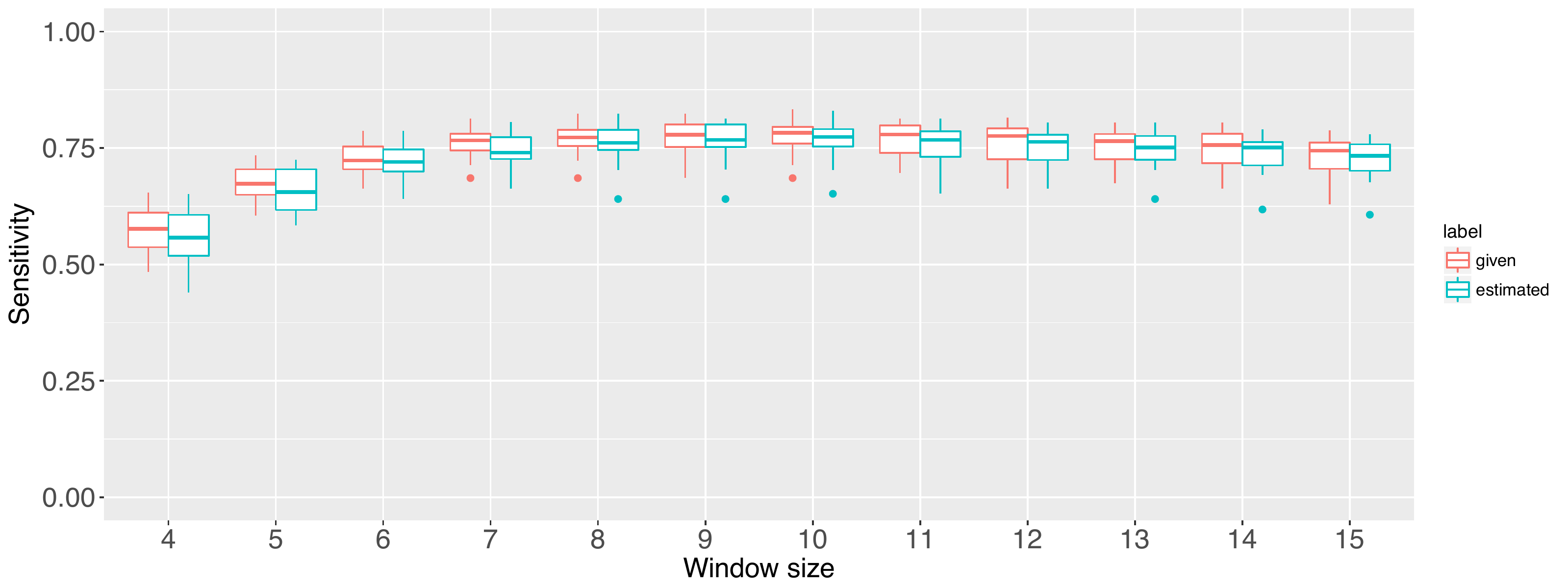}
  \caption{Sensitivity measures to evaluate the performance of the proposed method in graphs composed of 33 neurons sampled from a network of 100 neurons.
The horizontal axis represents the size of the time window: $\Delta=4, 5, 6, \cdots, 14, 15$.
For each $\Delta$, the left box plot represents the sensitivity estimated by Algorithm 2 given excitatory-inhibitory labels $\bm{z}$ and the right box plot represents the sensitivity estimated by Algorithm~2 and the estimated label $\bm{z}$.}
	\label{fig:sensitivity_100}
\end{figure}

\begin{figure}[h]
	\centering
	\includegraphics[bb=0 0 921 345, width=\linewidth]{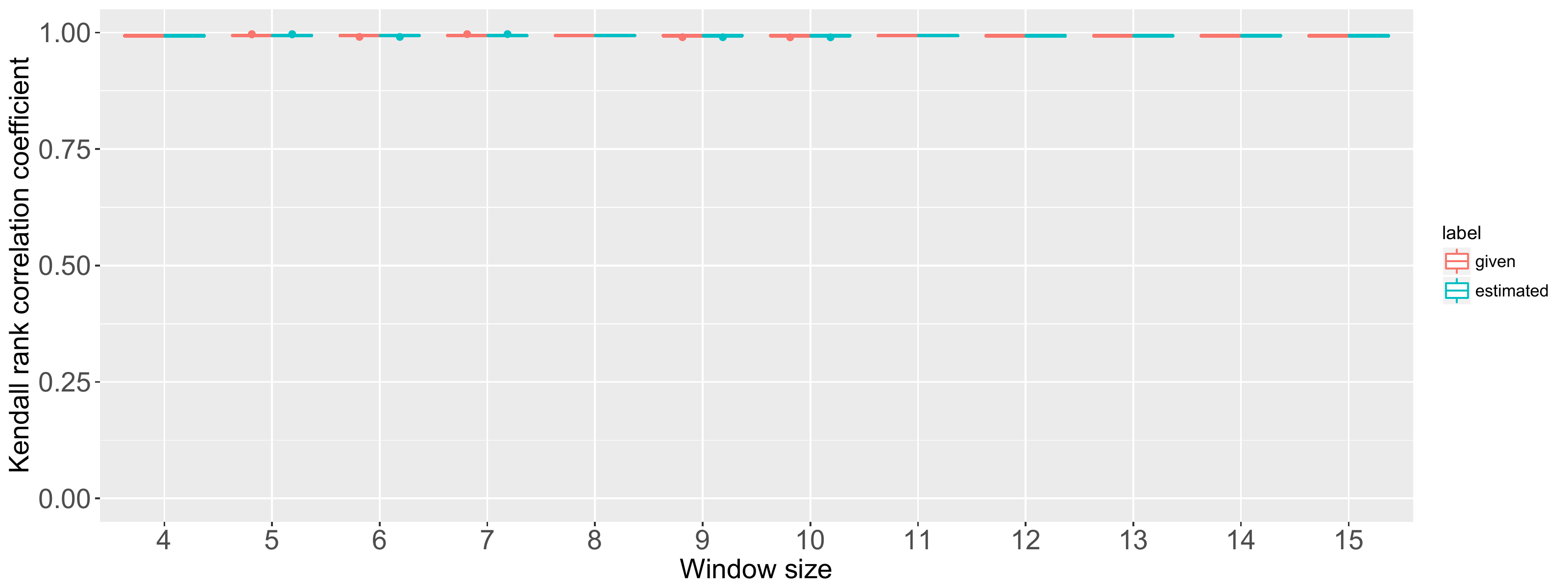}
	\caption{Kendall rank correlation coefficient measures to evaluate the performance of the proposed method in graphs composed of 33 neurons sampled from a network of 100 neurons.
The horizontal axis represents the size of the time window: $\Delta=4, 5, 6, \cdots, 14, 15$.
For each $\Delta$, the left box plot represents the Kendall rank correlation coefficient estimated by Algorithm 2 given excitatory-inhibitory label $\bm{z}$ and the right box plot represents the Kendall rank correlation coefficient estimated by Algorithm 2 and the estimated excitatory-inhibitory label $\bm{z}$.}
	\label{fig:kendall_100}
\end{figure}

\begin{figure}[H]
	\centering
	\includegraphics[bb=0 0 921 345, width=\linewidth]{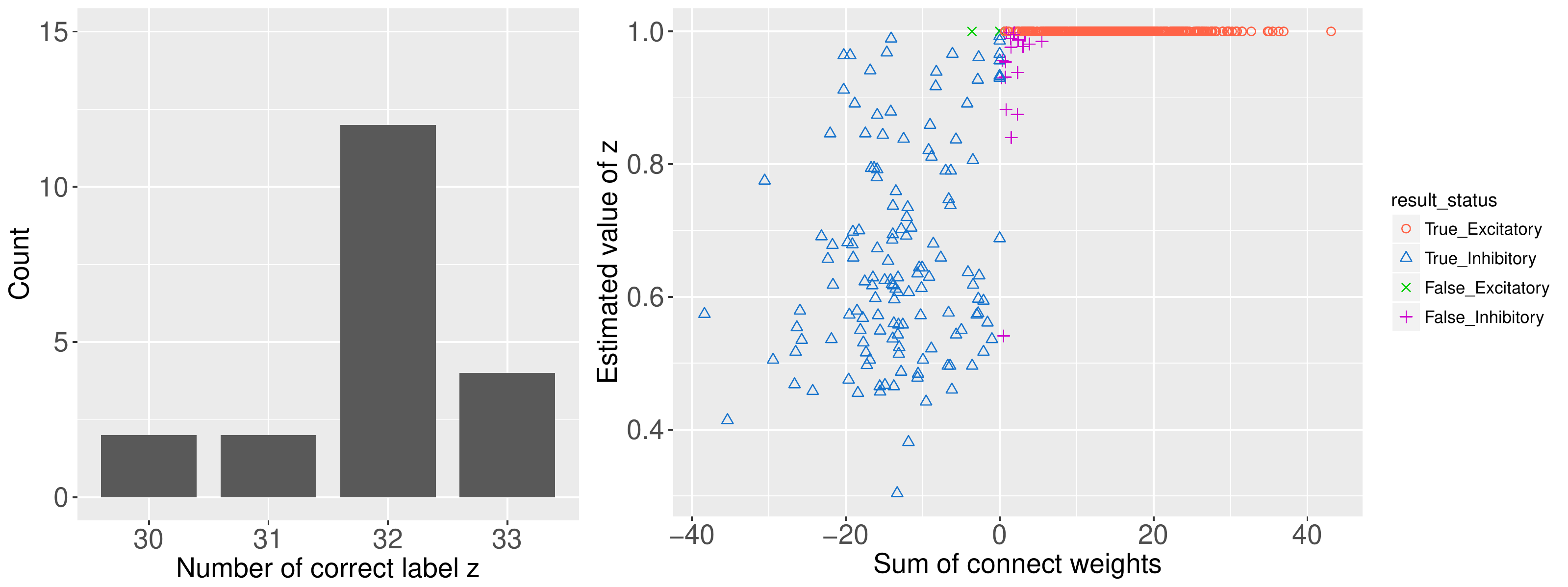}
  \caption{Estimation of label $\bm{z}$ for 33 neurons sampled from a network of 100 neurons.
  Left: Bar plot of the number of correctly estimated labels $\bm{z}$.
  Right: Estimated value of label $\bm{z}$ as a function of the sum of out-going connection weights of each neuron.
  The circles ($\circ$), plus signs ($+$), triangles ($\bigtriangleup$), and crosses ($\times$) represent correctly estimated excitatory neurons, incorrectly estimated excitatory neurons, correctly estimated inhibitory neurons, and incorrectly estimated inhibitory neurons, respectively.
  }
  \label{fig:scatter_alpha}
\end{figure}

We generated five different networks.
From each of them, 20 different graphs consisting of 33 neurons were sampled.
Starting from the window size $\Delta=4$, simulations were conducted until the sensitivity did not change significantly.
We examined 12 different time window sizes: $\Delta=4, 5, 6, \cdots, 14, 15$[ms].
To simplify the evaluation of the results, we removed weak connections from $W$ such that the number of connections in $W$ was the same as that in $W^*$.
\Figref{fig:sensitivity_100} and \Figref{fig:kendall_100} show the results of sensitivity and Kendall rank correlation coefficient analyses, respectively.
The horizontal axis represents the size of the time window, $\Delta=4, 5, 6, \cdots, 14, 15$.
For each value of $\Delta$, there are two box plots.
The ones on the left are the results obtained from Algorithm~2 given excitatory-inhibitory labels $\bm{z}$.
The ones on the right are the results obtained from Algorithm~2 and the estimated excitatory-inhibitory labels $\bm{z}$.
In \Figref{fig:sensitivity_100}, we see relatively high sensitivity, which peaks at around $0.76$ around $\Delta=10$[ms].
The results show that the proposed method estimated $76\%$ of $W^*$ connections correctly when spike interactions over $\sim 10$[ms] were considered.
When $\Delta > 10$[ms], i.e., in the case of a long time window, more propagations via multiple neurons were detected and the sensitivity was reduced.
In \Figref{fig:kendall_100}, the value of the Kendall rank correlation coefficient is nearly 1, suggesting that the proposed method detected the order of strong connections correctly.
Together with \Figref{fig:sensitivity_100}, we confirmed that our method detected the major connections in the network of 100 neurons successfully, even if only part of the neurons was observed.
Importantly, the performance did not decline significantly when the label $\bm{z}$ was also estimated from the data (shown in blue in \Figref{fig:sensitivity_100} and \Figref{fig:kendall_100}).
This is a particularly favorable property because the neuron type, whether excitatory or inhibitory, may not be easily obtained in real experiments.
To investigate how the proposed method with the estimated label $\bm{z}$ achieved the good performance, we show how the label $\bm{z}$ was estimated in the time window $\Delta=10$[ms] in \Figref{fig:scatter_alpha}.
On the left, the number of correctly estimated labels is plotted.
In \Figref{fig:scatter_alpha}, we see that the number of incorrectly estimated neurons is at most 3 among the 33 sampled neurons.
On the right, the estimated values of label $\bm{z}$ are plotted as a function of the sum of out-going connection weights of each neuron.
The results are summarized by four different symbols:~
correctly estimated excitatory neurons ($\circ$), incorrectly estimated excitatory neurons ($+$), correctly estimated inhibitory neurons ($\bigtriangleup$), and incorrectly estimated inhibitory neurons ($\times$).
We see that incorrectly estimated neurons have weak out-going connections, indicating that misclassification does not have a significant effect on the estimation of major connections.

\begin{figure}[h]
\begin{minipage}{0.5\hsize}
	\centering
	\includegraphics[bb=0 0 504 504, trim=50 75 50 0, width=\linewidth]{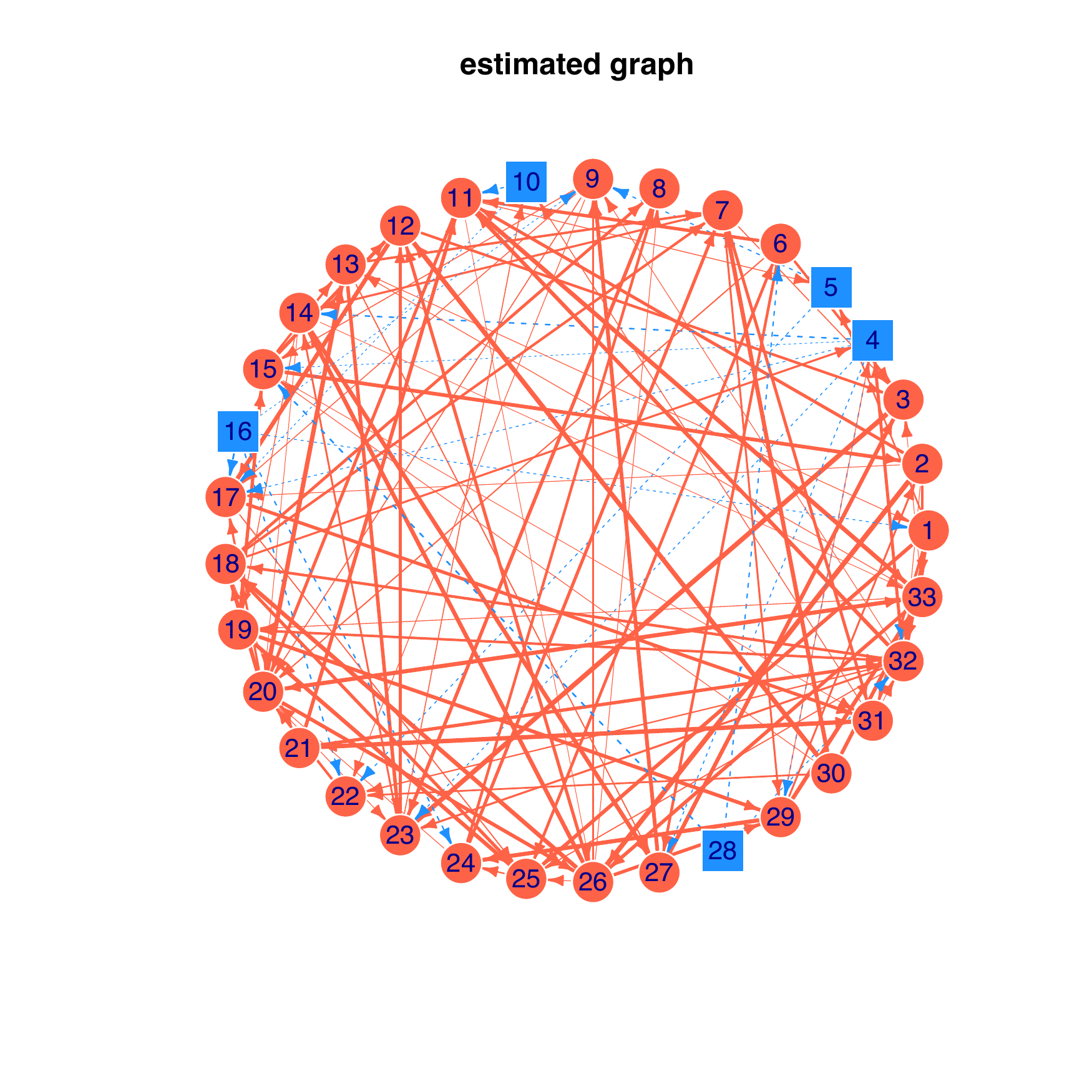}
\end{minipage}
\begin{minipage}{0.5\hsize}
	\centering
	\includegraphics[bb=0 0 504 504, trim=50 75 50 0, width=\linewidth]{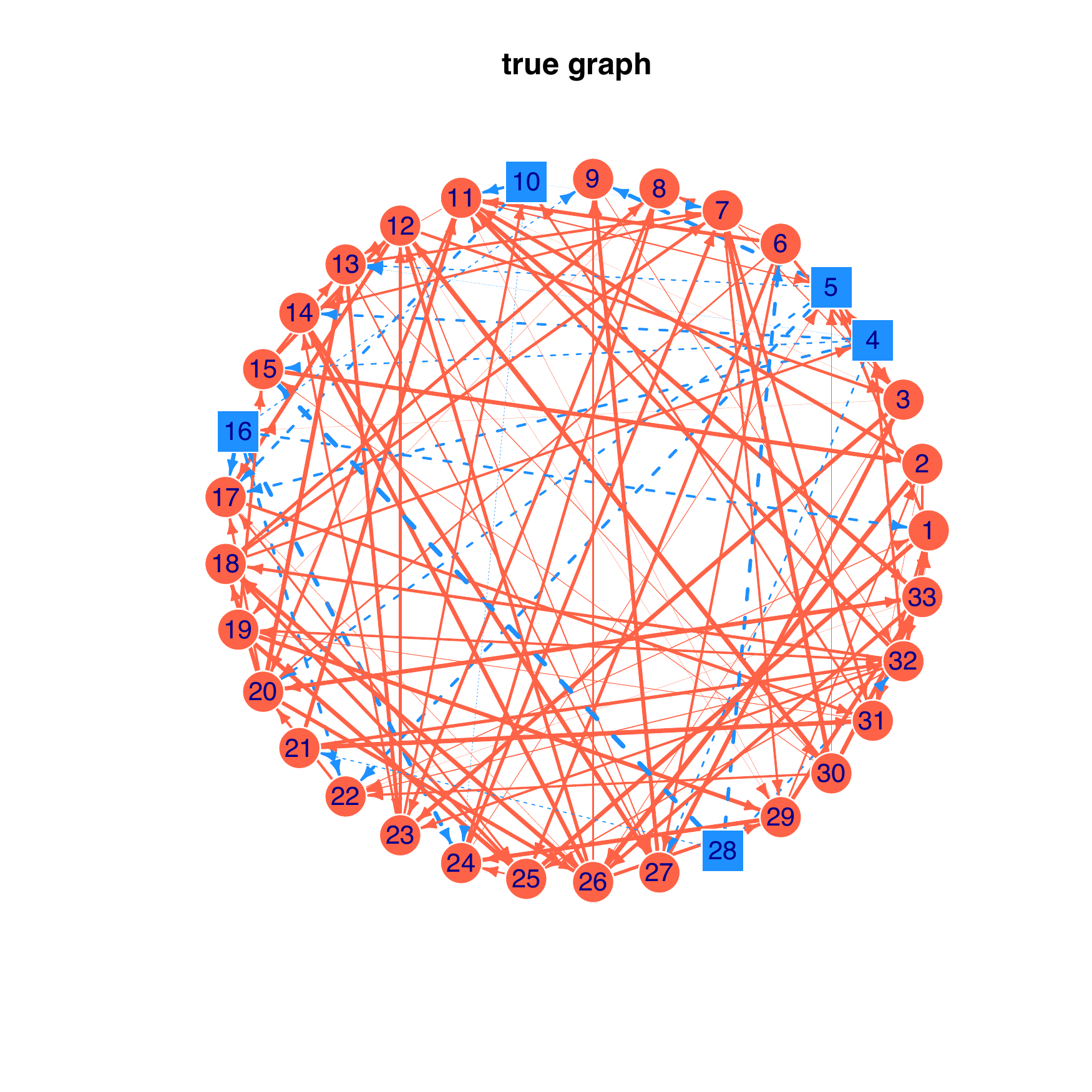}
\end{minipage}
  \caption{Example of the estimated graph $W$ and the true graph $W^*$.
  The red circular nodes represent excitatory neurons and the red arrows represent excitatory connections.
  The blue square nodes represent inhibitory neurons and the blue dashed arrows represent inhibitory connections.
  The thickness of the arrows represents the strength of connections.}
  \label{fig:graph_example}
\end{figure}

In \Figref{fig:graph_example}, we present an example of estimation.
The left graph shows the estimated connections $W$ and the right graph shows the true connections $W^*$.
The colors red and blue correspond to excitatory and inhibitory neurons, respectively, and the thickness of the edges represents the strength of the connections.
The results confirm that the proposed method can extract the major connections successfully.

\subsubsection{Network of 10,000 neurons}
\begin{figure}[h]
	\centering
	\includegraphics[bb=0 0 921 345, width=\linewidth]{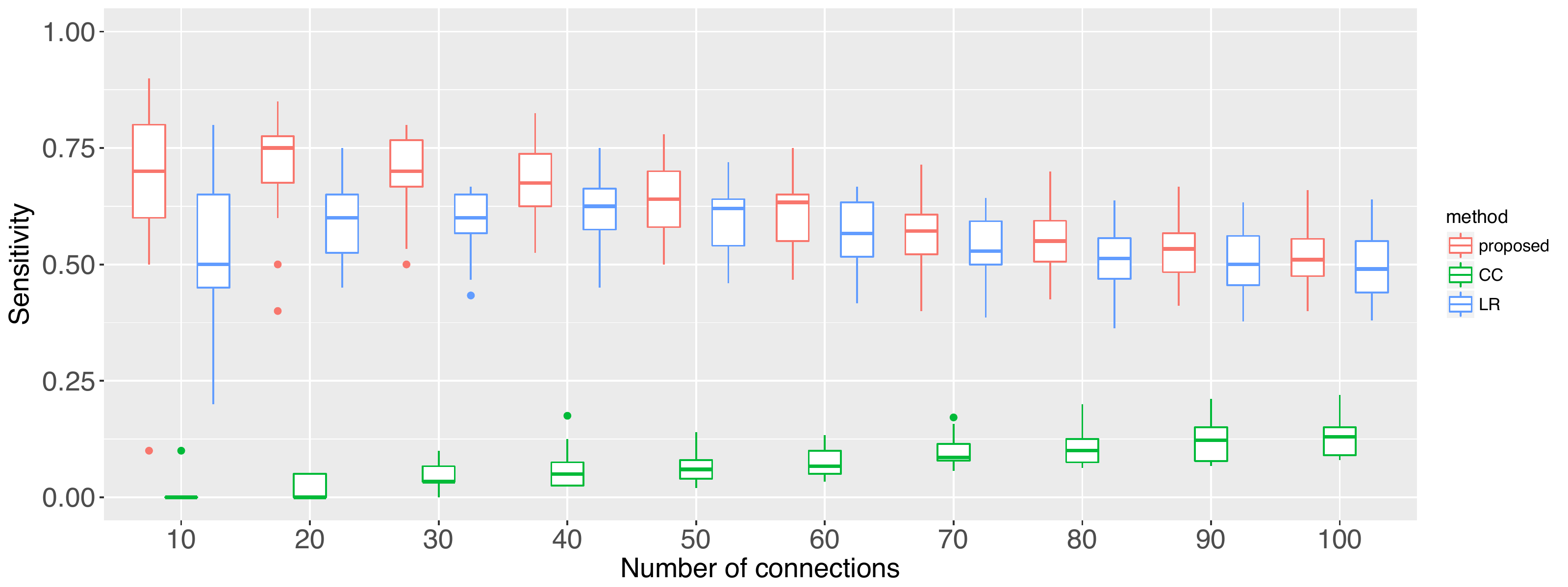}
	\caption{Sensitivity measures to evaluate the performance of the proposed method in graphs composed of 33 neurons sampled from a network of 10,000 neurons.
The horizontal axis represents the number of strongest connections from 10 to 100.
There are three box plots for each connection number (from left to right, the results by Algorithm~2 and the estimated excitatory-inhibitory labels $\bm{z}$, the cross-correlation function, and the binary logistic regression).}
	\label{fig:sens_10000}
\end{figure}

\begin{figure}[H]
	\centering
	\includegraphics[bb=0 0 921 345, width=\linewidth]{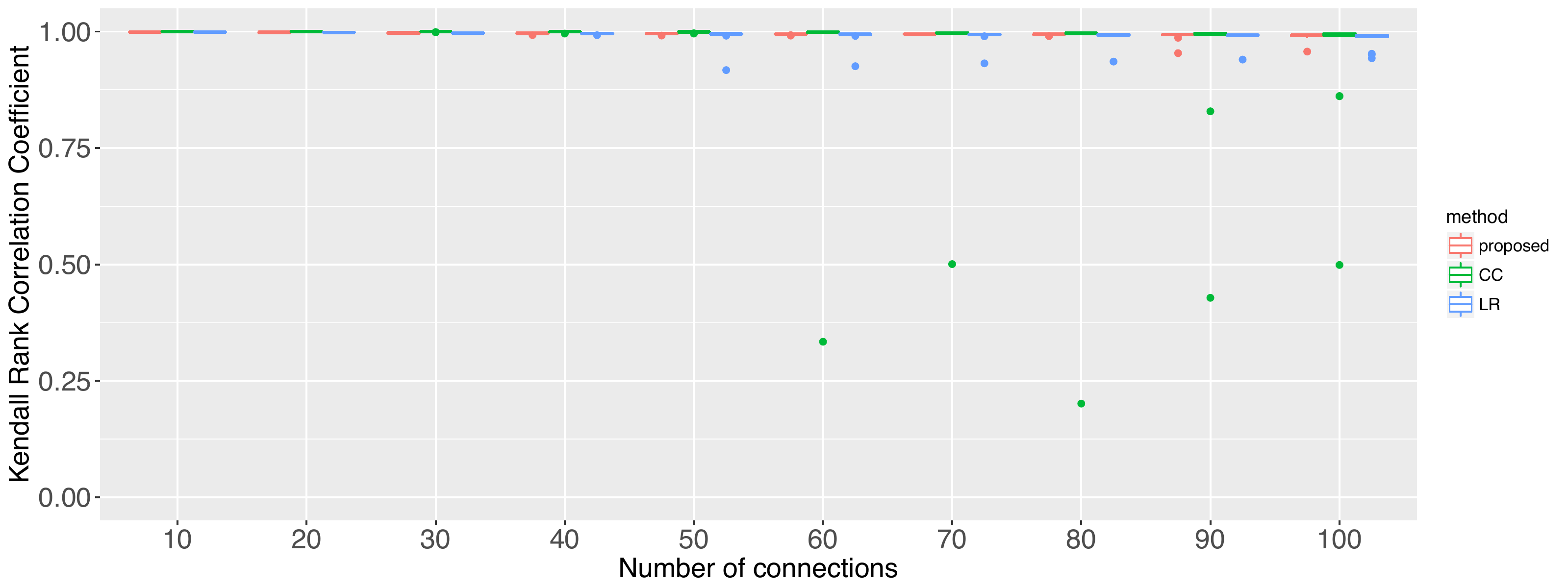}
	\caption{Kendall rank correlation coefficient measures to evaluate the performance of the proposed method in graphs composed of 33 neurons sampled from a network of 10,000 neurons.
The horizontal axis represents the number of strongest connections from 10 to 100.
There are three box plots for each connection number
(from left to right, the results by Algorithm~2 and the estimated excitatory-inhibitory labels $\bm{z}$, the cross-correlation function, and the binary logistic regression.}
	\label{fig:kend_10000}
\end{figure}

We generated one network.
From this network, 20 different graphs consisting of 33 neurons were sampled.
This simulation was performed to verify the proposed method under conditions comparable to real neuroscience experiments.
Motivated by the result of the 100-neuron network, we set the size of the time window to $\Delta=10$[ms].
We removed small values of the estimated connections $W$ and set the number of connections in $W$ to a pre-set value.
Further, we assessed the performances by the correct detection of the strongest connections between observed neuron pairs.
We systematically varied the number from the top $10$ connections to the top $100$ connections.
The sensitivity and Kendall rank correlation coefficient are plotted in \Figref{fig:sens_10000} and \Figref{fig:kend_10000}, respectively, as functions of the number of connections.
There are three box plots for each number of connections (from left to right, the results by Algorithm~2 and the estimated excitatory-inhibitory labels $\bm{z}$, the cross-correlation function, and the binary logistic regression).
In \Figref{fig:sens_10000}, we observe that the proposed method outperforms the other methods.
In particular, on average, our method extracted $70\%$ of the connections correctly, up to the top $20\%$ of the connections.
In terms of the Kendall rank correlation coefficient, as shown in \Figref{fig:kend_10000}, these three methods estimate the rank of connections nearly equally well.

\begin{figure}[H]
	\centering
	\includegraphics[bb=0 0 921 345, width=\linewidth]{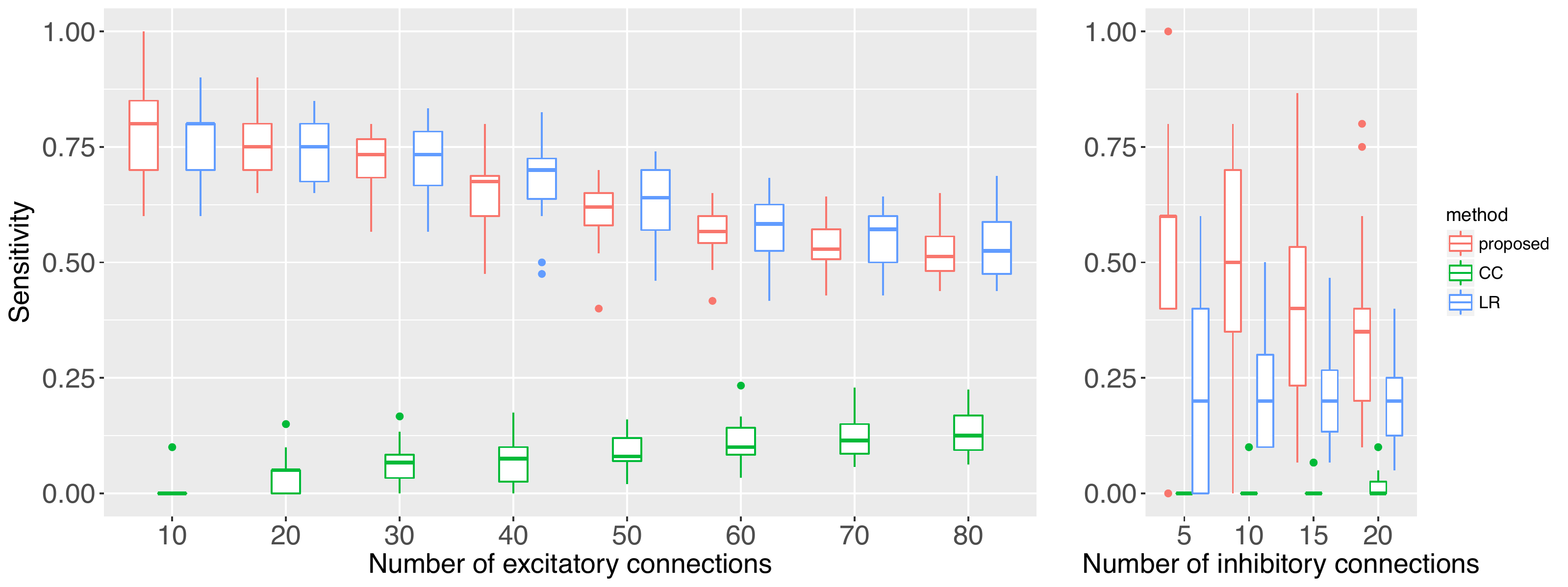}
  \caption{Sensitivity evaluated separately for excitatory and inhibitory connections in graphs composed of 33 neurons sampled from a network of 10,000 neurons.
  Left:~the horizontal axis represents the number of strongest excitatory connections from 10 to 80.
  Right:~the horizontal axis represents the number of strongest inhibitory connections from 5 to 20.
  There are three box plots for each connection number
(from left to right, the results by Algorithm~2 and the estimated excitatory-inhibitory labels $\bm{z}$, the cross-correlation function, and the binary logistic regression.
  }
  \label{fig:sens_10000_exc_inh}
\end{figure}

\begin{figure}[H]
	\centering
	\includegraphics[bb=0 0 921 345, width=\linewidth]{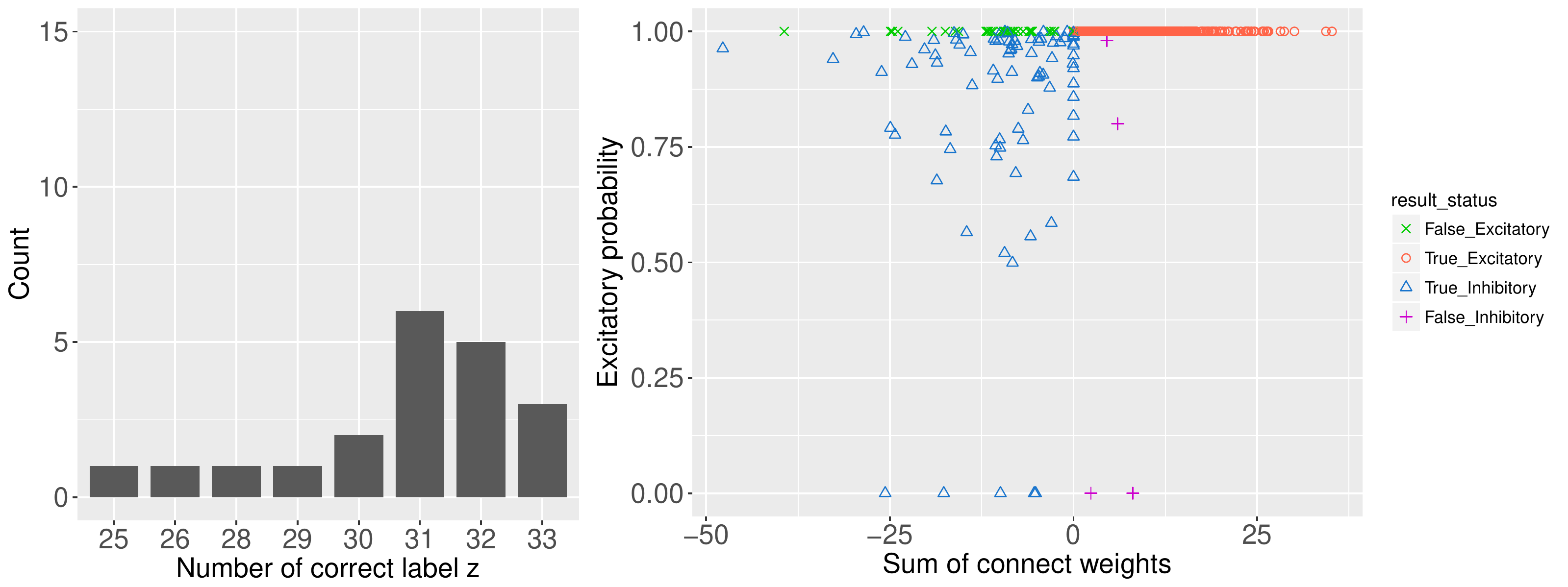}
  \caption{Estimation of label $\bm{z}$ for 33 neurons sampled from a network of 10,000 neurons.
  Left: bar plot of the number of correctly estimated labels $\bm{z}$.
  Right: estimated value of label $\bm{z}$ as a function of the sum of out-going connection weights of each neuron.
  The circles ($\circ$), plus signs, ($+$), triangles ($\bigtriangleup$), and crosses ($\times$) represent correctly estimated excitatory neurons, incorrectly estimated excitatory neurons, correctly estimated inhibitory neurons, and incorrectly estimated inhibitory neurons, respectively.
  }
	\label{fig:scatter_alpha_10000}
\end{figure}

We also investigated the sensitivity for excitatory and inhibitory connections separately.
As shown in the left column of \Figref{fig:sens_10000_exc_inh}, the proposed method as well as logistic regression exhibited high sensitivity for excitatory connections.
Specifically, the average sensitivity of the proposed method for the top 20 connections was greater than $75\%$.
\citet{song2005highly} demonstrated that the top $17\%$ of synaptic connections contribute to half of the total connection strength.
The result suggests that the proposed method detects the majority of the major excitatory connections successfully.
Regarding inhibitory connections, as shown in the right column of \Figref{fig:sens_10000_exc_inh}, the proposed method clearly outperformed the other two methods.
Such excellent performance is due to two features of the proposed method: removal of nuisance inputs and decomposition of pseudo-connections.
In addition, we investigated how the proposed method estimates excitatory-inhibitory labels $\bm{z}$ in \Figref{fig:scatter_alpha_10000}, in the same way as in \Figref{fig:scatter_alpha}.
We found that many neuron types are estimated correctly in \Figref{fig:scatter_alpha_10000} left.
We also confirmed that nearly all excitatory neurons are estimated correctly
\Figref{fig:scatter_alpha_10000} right.

From an overall perspective, the numerical investigation using partially observed synthetic data demonstrated that the proposed method can detect both excitatory and inhibitory connections reliably.

\section{Real data experiment}
In this section, we apply the proposed method to electrophysiological data.
We investigate whether the behaviorally induced effective connectivity assessed by the proposed method is sustained during a post-task sleep epoch.

\subsection{Experiment outline}
The electrophysiological data analyzed in this study were originally reported in \citet{Tatsuno2006methodological}.
Briefly, an adult male brown Norway/Fischer 344 hybrid rat was used for 25 h of continuous recording.
Based on the experimental protocol of \citet{ribeiro2004long}, the recording sessions consisted of three epochs: a 12-h free-running pre-task epoch, a 1-h task epoch, and a 12-h post-task epoch.
After implantation of the microdrive, the rat was housed in a recording box (height 42[cm], length 46.5[cm], and width 46.5[cm]) for at least 1 week before recording.
This ensured that the animal was accustomed to the recording environment.
Throughout the recording, the rat was allowed to move, eat, and sleep freely in the recording box, following its preferred sleep/wake cycle.
During the task, the animal explored four novel objects located at each corner of the recording box.
Encounters with these novel objects were considered novel experiences, as in a study by \citet{ribeiro2004long}.

The recording was made by the microdrive with 12 independently adjustable tetrodes, covering a circular area of approximately 1 mm in diameter \citep{gothard1996dynamics}.
The drive was implanted above the hippocampus [3.8 mm posterior and 2.5 lateral (left) to the bregma], and lowered to the CA1 area.
A reference electrode was implanted in the corpus callosum and an electrode for recording the hippocampal theta oscillation was implanted in the hippocampal fissure.
The neural signals were bandpass filtered between 600 Hz and 6 kHz, and spike waveforms were recorded at 32 kHz whenever the signal exceeded a predetermined threshold.
The recording of all data was performed using Cheetah Data Acquisition Systems (Neuralynx, Bozeman, MT, USA).
The rat's head position was identified by light emitting diodes on the microdrive and monitored by a color camera mounted on the ceiling of the recording room.
The rat was also monitored by an infrared camera to allow for observation of behavior during the dark cycle.
The video data were time-stamped and used for off-line detection of motion and motionless periods.
To identify REM episodes within motionless periods, local field potential (LFP) traces were bandpass filtered in the delta (2--4 Hz) and theta (6--10 Hz) bands.
The power in each band was computed and REM episodes were identified as periods of elevated theta-delta power ratio, such as $> 2.0$.
The rest of the motionless periods was considered non-REM sleep.
The units were isolated using a multidimensional cluster cutting software (MClust by A. D. Redish, University of Minnesota, Minneapolis, MN, USA).
Only units with $<1\%$ of inter-spike-interval distribution falling within the 2 ms refractory period were used in the analysis.
This process yielded 48 CA1 units.
The type of neuron (excitatory or inhibitory) was estimated using the firing rate and spike width \citep{markus1994spatial}.
For detection of effective connectivity by the proposed method, neurons with a low firing rate ($< 0.5$ Hz) were excluded because such neurons prevent reliable detection of connectivity.

\subsection{Data analysis}
\subsubsection{Target data of interest}
\begin{figure}[h]
	\centering
  \includegraphics[bb=0 0 702 237, width=\linewidth]{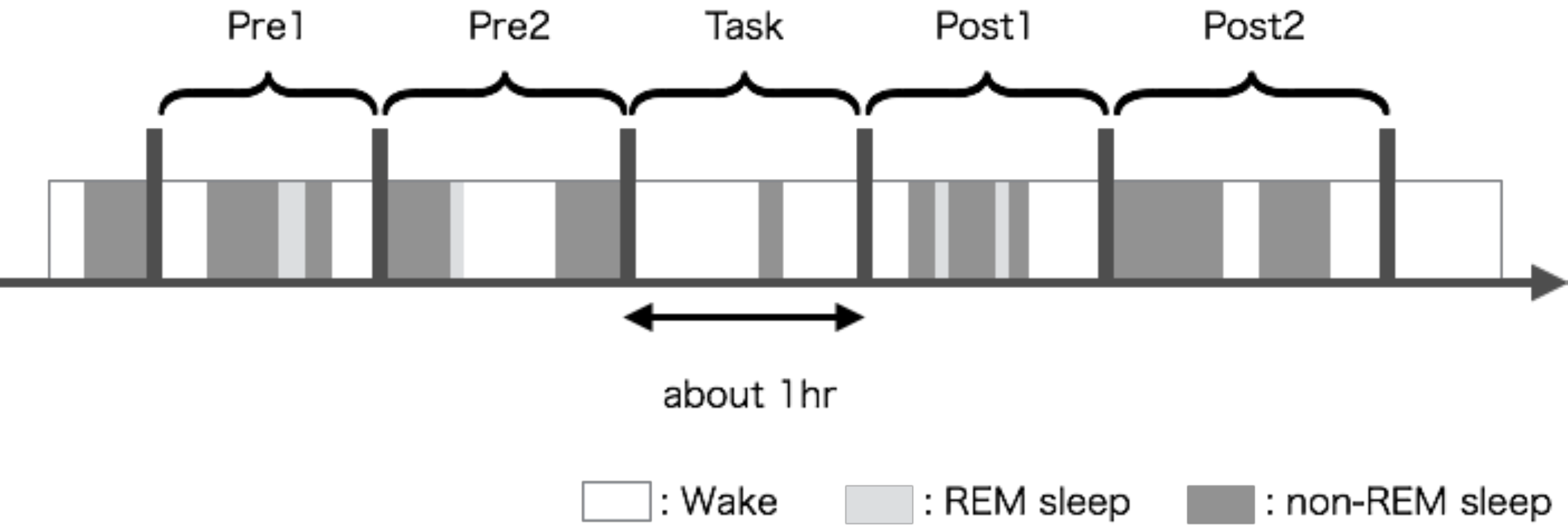}
  \caption{Schematic illustration of electrophysiological data.
  The 5 h of spike data were divided into 5 segments of nearly equal duration:~Pre1, Pre2, Task, Post1, and Post2.
  The colors white, light gray, and dark gray represent wake, REM sleep, and non-REM sleep epochs, respectively.
  Waking portions during the task epoch and non-REM portions during the pre-task and post-task epochs were used in the analysis.
  }
  \label{fig:realdata}
\end{figure}

We focused on 5 h of recording data (2 h immediately before the task, 1 h of the task, and 2 hours immediately after the task) because memory replay of this type of spatial memory task is typically detected within 1 h after the task experience \citep{kudrimoti1999reactivation, Tatsuno2006methodological}.
For analytical clarity, 2-h pre-task and post-task epochs were divided into approximately 1-h segments, respectively, which yielded 5 diagrams, as shown in \Figref{fig:realdata}.
Following \citet{wilson1994reactivation}, we analyzed the non-REM episodes during the pre-task and post-task epochs and the waking activity during task epoch.

\subsubsection{Results}
\begin{figure}[H]
  \centering
  \includegraphics[bb=0 0 609 1051, trim=0 0 10 10, clip, width=.75\linewidth]{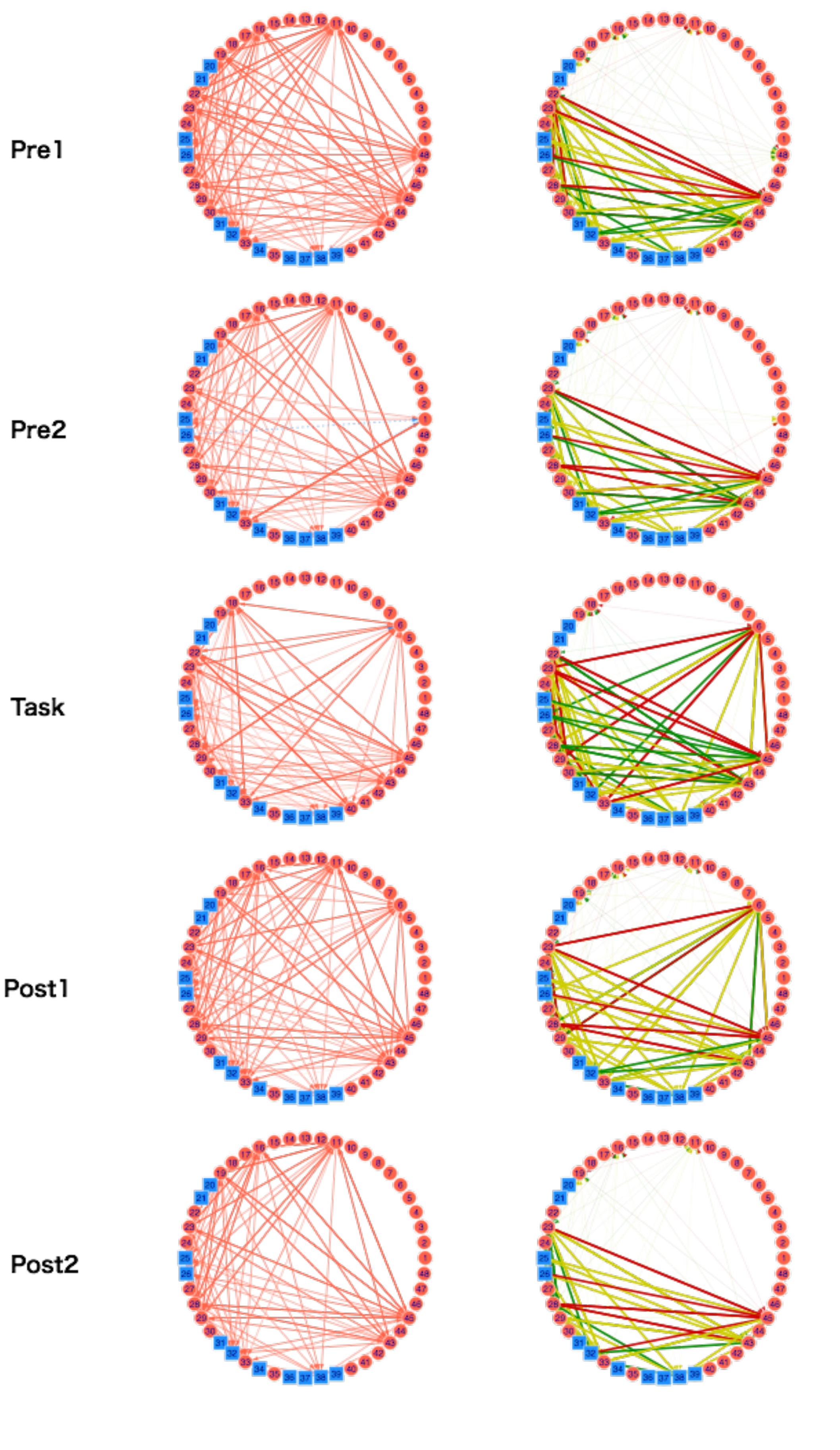}
  \caption{Estimated effective connectivity for each epoch.
  From top to bottom, each diagram represents the Pre1, Pre2, Task, Post1, and Post2 epoch, respectively.
  Left column: estimated effective connectivity drawn as in \Figref{fig:graph_example}.
  Right column: the estimated effective connectivity drawn as by \citet{wilson1994reactivation}.
  The red, green, and yellow lines represent connection weights of the top $25\%$, $50\%$, and $100\%$, respectively.
  The bold lines represent the connections that exist during the task and during either the pre-task or the post-task sleep epochs (Pre1, Pre2, Post1, and Post2).
  }
  \label{fig:graphs_real}
\end{figure}

The size of the time window of the proposed method was set to $15$ ms on the basis of the spike-timing-dependent plasticity (STDP) window \citep{bi1998synaptic}.
Because the excitatory and inhibitory neurons were already estimated from the spike waveforms, we used the proposed method under the condition that the excitatory-inhibitory labels were already known.
The estimated effective connectivity during the five segments is shown in the left column of \Figref{fig:graphs_real}.
To quantify the similarity between waking and sleep epochs in the connection diagram, we defined the overlap rate $o^{(p)}$ for estimated connections $\{w_{ij}\}_{i,j\in V}$ by
\begin{align}
  o^{(p)} = \frac{\sum_{i,j}\Bigl(|{\rm sgn}(w_{ij})| \cdot |{\rm sgn}\bigl(\tilde{w}^{(p)}_{ij}\bigr)|\Bigr)}{\sum_{i,j} |{\rm sgn}\bigl(\tilde{w}^{(p)}_{ij}\bigr)|},
\end{align}
where ${\rm sgn}(\cdot)$ is the sign function.
The top $p\%$ connections $\{\tilde{w}_{ij}^{(p)}\}_{i,j\in V}$ during the task epoch were obtained by truncating the other weaker connections:
\begin{align}
  \tilde{w}^{(p)}_{ij} = \begin{cases}
    \tilde{w}_{ij} & \tilde{w}_{ij}>\zeta^{(p)}, \\
    0 & \mbox{otherwise},
\end{cases}\nonumber
\end{align}
where $\zeta^{(p)}$ is the strength of
the top $p\%$ connections.
Table \ref{tb:overlap_task} summarizes the overlap between the estimated connections during the task and during sleep epochs (Pre1, Pre2, Post1 and Post2).
For the top $25\%$ connections, the overlap rate of Post1 is the highest among the four sleep epochs; nearly half of the behaviorally induced connections remain during Post1.
For the top $50\%$ and $100\%$, the Post1 similarity is weakened, suggesting that the learned information is encoded by strong connections.
The right column of \Figref{fig:graphs_real} shows that new effective connections emerge on top of the existing stationary connections, and they are sustained during Post1.
The result is also consistent with the previous finding that memory replay is typically observed within 30 min after the task training~\citep{kudrimoti1999reactivation}.

\begin{table}[h]
	\centering
  \caption{Overlap rate of estimated connections between the task and sleep epochs.}
	\begin{tabular}{c||c|c|c|c}
    \hline
    $p\%$ & Pre1 & Pre2 & Post1 & Post2 \\
    \hline
    $25\%$ & 0.286 & 0.200 & {\bf 0.429} & 0.229 \\
    $50\%$ & 0.486 & 0.357 & 0.429 & 0.300 \\
    $100\%$ & 0.471 & 0.364 & 0.450 & 0.336 \\
  \hline
	\end{tabular}
	\label{tb:overlap_task}
\end{table}

\section{Conclusion}
In this paper, we proposed a novel method to estimate neural connections from partially observed spike trains.
Specifically, we addressed three mathematical difficulties:~influence of pseudo-correlations, influence from unobserved neurons, and estimation of excitatory and inhibitory neurons (hence, excitatory and inhibitory directed connections).
The proposed method extracted the connections between the observed neurons by explicitly quantifying the influence from unobserved neurons.
This was achieved by using the property of a sum of random variables and the assumption that nuisance inputs follow Gaussian distributions.
Then, by modeling all possible spike propagations over multiple neurons, we eliminated the pseudo-correlations.
Finally, we proposed a framework for estimating neuron types (excitatory or inhibitory) using the $em$ algorithm.
Through numerical simulations, we confirmed that the proposed method can detect the major connection reliably for a small random network as well as for a large realistic network.
In particular, for the estimation of inhibitory connections, the proposed method outperformed other standard methods, such as the cross-correlation function and binary logistic regression.
In addition, we applied the method to multi-electrode data from the CA1 area of rat's hippocampus.
We found that the behavior-induced effective connections and the detected connections were sustained during the subsequent sleep.

The proposed method can be used for providing a macroscopic measure of connections within a temporal window of interest and for selecting specific connections for further analyses.
However, note that it estimates averaged connections within a certain observation period specified by the width of the temporal window, but not dynamical connections.
The validity of the average of neural connections is discussed in~\citep{toyoizumi2009mean}.

There are some issues that warrant further investigation.
First, the proposed method tends to estimate inhibitory connections that are weaker than excitatory connections.
This problem may arise from the same activation functions for both excitatory and inhibitory neurons even though they tend to fire at different frequencies.
Separate activation functions may alleviate this problem.
Second, in the present approach, the number of connections is given as an ad hoc parameter to the proposed algorithm.
In other words, the proposed method does not have a mechanism to estimate the number of connections from each neuron.
This could be determined by stability-based methods, such as StARS~\citep{liu2010stability} and TIGRESS~\citep{haury2012tigress}, but it should be noted that these methods tend to overestimate the number of connections.
An important direction for future work is to develop an appropriate method to estimate the number of connections by considering the assumed signal propagation model.

\section*{Acknowledgements}
Part of this work was supported by JSPS KAKENHI No.15H02669, 16K16108, 17H01793, and 25120009, and JST CREST ACA20935 and MJCR14D7.

\appendix
\section{Proof of the property of the sum of random variables}
\subsection{Proof of Theorem 1}
Let $X$ and $Y$ be independent random variables with probability density functions $f_X$ and $f_Y$, respectively.
We show the proof of the property of the sum $X+Y$ when $\mathbb{E}[Y]=0$.
Consider any bonded function $g$.
The expectation of $g(X+Y)$ is represented by
\begin{align}
	\mathbb{E}[g(X+Y)] &= \int g(z)f_{X+Y}(z)dz\nonumber\\
	&= \int g(z) \Biggl[\int f_{X}(x)f_{Y}(z-x)dx \Biggr] dz\nonumber\\
	&= \iint f_{X}(x) g(z) f_{Y}(z-x) dzdx.
\end{align}
We define $f^{-}_{Y}(x) = f_{Y}(-x)$ and obtain
\begin{align}
	\iint f_{X}(x) g(z)f_{Y}(z-x)dzdx
	&= \int f_{X}(x)\Biggl(\int g(z)f^{-}_{Y}(x-z)dz\Biggr)dx
	\nonumber\\
	&= \int f_X(x)h(x)dx\nonumber\\
	&= {\mathbb E}[h(X)],
\end{align}
where $h=g*f^{-}_{Y}$, i.e., convolution of functions $g$ and $f^{-}_{Y}$.
Thus, we get the following equation:
\begin{align}
	{\mathbb E}[g(X+Y)]
	= {\mathbb E}[h(X)].
	\label{eq:var_sum}
\end{align}
Then, we show the proof when $\mathbb{E}[Y]\neq0$.
In this case, we transform variables $X$ and $Y$ into $X'$ and $Y'$ as follows:
\begin{align}
	X'&=X+\mathbb{E}[Y],
	\nonumber\\
	Y'&=Y-\mathbb{E}[Y].
	\nonumber
\end{align}
As with the proof when $\mathbb{E}[Y]=0$, we obtain
\begin{align}
	{\mathbb E}[g(X+Y)]
	&= {\mathbb E}\Bigl[g\Bigl((X+\mathbb{E}[Y])+(Y-\mathbb{E}[Y])\Bigr)\Bigr]
	\nonumber\\
	&= {\mathbb E}[g(X'+Y')]
	\nonumber\\
	&= {\mathbb E}[h(X')].
\end{align}
Thus, we obtain the following equation:
\begin{align}
	{\mathbb E}[g(X+Y)]
	= {\mathbb E}[h(X+\mathbb{E}[Y])].
	\label{eq:var_sum}
\end{align}

\subsection{Proof of Corollary~1}
Assume that a function $g$ is Gaussian density function $\phi_{\sigma^2}(x)=\int_{-\infty}^{x}\mathcal{N}(z; 0, \sigma^2)dz$ and $f_Y$ is a Gaussian distribution with mean $\mathbb{E}[Y]$ and variance $\tau^2$.
The Gaussian density function is symmetric with $\mathbb{E}[Y]$; hence, $f^{-}_Y = f_Y$.
The convolution of $g$ and $f^{-}_Y$ is represented by
\begin{align}
	g*f^{-}_Y = \phi_{\sigma^2} * \phi_{\tau^2} = \phi_{\sigma^2 + \tau^2},
\end{align}
and we obtain
\begin{align}
	\mathbb{E}[\phi_{\sigma^2}(X+Y)] = \phi_{\sigma^2+\tau^2}(X+\mathbb{E}[Y]).
\end{align}

\subsection{Proof of Theorem 2}
Consider a cumulative distribution function $G$ of a probability density function $g$.~
First, we show the following equation:
\begin{align}
	G*f(x) = \int_{-\infty}^{x} g*f(z) dz,
	\label{eq:cum_conv}
\end{align}
where $f$ is a probability density function.
The right-hand side of the above equation is transformed as follows:
\begin{align}
	G*f(x) &= \int G(y)f(x-y)dy
	\nonumber\\
	&= \iint g(z)\mathbb{1}_y(z)f(x-y)dz\;dy,
\end{align}
where $\mathbb{1}_y(z), z\in\mathbb{R}$ is the following function:
\begin{align}
	\mathbb{1}_y(z) =
	\left\{
	\begin{aligned}
		&1,\qquad& z<y,\\
		&0,\qquad& z>y.
	\end{aligned}
	\right.
\end{align}
We define $w=x-y$ and obtain
\begin{align}
	\iint g(z)\mathbb{1}_y(z)f(x-y)dz\;dy
	&= \iint g(z)\mathbb{1}_{x-w}(z)f(w)dz\;dw\qquad z\in[-\infty, x-w]
	\nonumber\\
	&= \int_{-\infty}^{\infty}\int_{-\infty}^{x-w} g(z)f(w)dz\;dw.
	\label{eq:left_cum_conv}
\end{align}
On the other hand, the left-hand side of Eq. \ref{eq:cum_conv} is transformed as follows~:
\begin{align}
	\int_{-\infty}^{x} g*f(z) dz
	&= \int_{-\infty}^{x}\int_{-\infty}^{\infty}g(z)f(y-z)dz\;dy
	\nonumber\\
	&= \iint g(z)f(y-z)\mathbb{1}_x(y)dz\;dy.
\end{align}
We define $w=y-z$ and obtain
\begin{align}
	\iint g(z)f(y-z)\mathbb{1}_x(y)dz\;dy
	&= \iint g(z)f(w)\mathbb{1}_x(w+z)dz\;dw
	\nonumber\\
	&= \int_{-\infty}^{\infty}\int_{-\infty}^{x-w} g(z)f(w)dz\;dw.
	\label{eq:right_cum_conv}
\end{align}
Eq. \ref{eq:left_cum_conv} coincides with Eq. \ref{eq:right_cum_conv} and it proves Eq. \ref{eq:cum_conv}.
By applying Theorem 1 to the expectation of $G(X+Y)$, we get the following equation~:
\begin{align}
	\mathbb{E}[G(X+Y)] &= \mathbb{E}[H(X+\mathbb{E}[Y])],
\end{align}
where $H(x) = (G*f^{-}_{Y})(x) = \int_{-\infty}^{x}(g*f^{-}_{Y})(z)dz$.

\subsection{Proof of Corollary~2}
Consider the case $G(x)=\Phi_{\sigma^2}(x)=\int \phi_{\sigma^2}(z)dz$ and let $f_Y$ be a Gaussian distribution with mean $\mathbb{E}[Y]$ and variance $\tau^2$.
The convolution of $G$ and $f^{-}_Y$ is represented by
\begin{align}
	(G*f^{-}_Y)(x) &= \int_{-\infty}^{x} (g*f^{-}_Y)(z) dz
	\nonumber\\
	&= \int_{-\infty}^{x} (\phi_{\sigma^2} * \phi_{\tau^2})(z) dz
	\nonumber\\
	&= \int_{-\infty}^{x} \phi_{\sigma^2 + \tau^2}(z) dz
	\nonumber\\
	&= \Phi_{\sigma^2 + \tau^2}(x).
\end{align}
Therefore, we obtain
\begin{align}
	\mathbb{E}[\Phi_{\sigma^2}(X+Y)] = \Phi_{\sigma^2+\tau^2}(X+\mathbb{E}[Y]).
\end{align}

\section{Derivation of projection steps in the $em$ algorithm to estimate excitatory probability}
\subsection{Derivation of the projection from $\mathcal{D}$ to $\mathcal{D}'$}
Consider the $e$-projection from $\mathcal{D}$ to $\mathcal{D}'$:
\begin{align}
  \bm{\beta}^{(\tau+1)} = \argmin_{\bm \beta} D_{\mathrm{KL}}[q'(\Lambda, \bm{z}; \bm{\beta}), q(\Lambda, \bm{z}; \bm{\alpha}^{(\tau)})].
\end{align}
Substituting $p(\Lambda, \bm{z}; \bm{\alpha}^{(\tau)})$ with $q(\Lambda, \bm{z}; \bm{\alpha}^{(\tau)})$ in the previous projection step, the above equation is equivalent to
\begin{align}
	\bm{\beta}^{(\tau+1)} &= \argmin_{\bm \beta} D_{\mathrm{KL}}[q'(\bm{z} \mid \Lambda; \bm{\beta}), p(\bm{z}\mid\Lambda; \bm{\alpha}^{(\tau)})]
  \nonumber\\
	&=\argmin_{\bm \beta} \sum_{\bm{z}} q'(\bm{z}\mid\Lambda; \bm{\beta})\log{q'(\bm{z}\mid\Lambda; \bm{\beta})}
	- q'(\bm{z}\mid\Lambda; \bm{\beta})\log{p(\bm{z}\mid\Lambda; \bm{\alpha}^{(\tau)})}
  \nonumber\\
	&=\argmin_{\bm \beta}\sum_{\bm{z}}\Bigl\{
    \prod_{i=1}^{N}\beta_i^{z_i}(1-\beta_i)^{1-z_i}
    \log{\prod_{i=1}^{N}\beta_i^{z_i}(1-\beta_i)^{1-z_i}}
	  \nonumber\\
    &\qquad\qquad\qquad
    -\prod_{i=1}^{N}\beta_i^{z_i}(1-\beta_i)^{1-z_i}
    \log{p(\bm{z}\mid\Lambda; \bm{\alpha}^{(\tau)})}
    \Bigr\}
  \nonumber\\
	&=\argmin_{\bm \beta}\sum_{\bm{z}}\Biggl[
  \prod_{i=1}^{N}\beta_i^{z_i}(1-\beta_i)^{1-z_i}
	\Biggl\{
    \log{\prod_{i=1}^{N}\beta_i^{z_i}(1-\beta_i)^{1-z_i}}
	  -\log{p(\bm{z}\mid\Lambda;\bm{\alpha}^{(\tau)})}
  \Biggr\}\Biggr].
  \label{eq:argmin_beta}
\end{align}
Here, we define
\begin{align}
	q'(\bm{z}_{-i}\mid\Lambda, \bm{\beta}) = \prod_{j \neq i}\beta_j^{z_j}(1-\beta_j)^{1-z_j}\nonumber
\end{align}
and partially differentiate the terms in Eq. \ref{eq:argmin_beta} with the parameter $\beta_i$.
\begin{align}
	&\frac{\partial}{\partial \beta_i}\prod_{i=1}^{N}\beta_i^{z_i}(1-\beta_i)^{1-z_i}
	= q'(\bm{z}_{-i}\mid\Lambda, \bm{\beta})\Biggl\{z_i
    \Bigl(
      \frac{1-\beta_i}{\beta_i}
    \Bigr)^{1-z_i}
    - (1-z_i)\Bigl(
      \frac{\beta_i}{1-\beta_i}\Bigr)^{z_i}
      \Biggr\}
  \nonumber\\
	&\frac{\partial}{\partial \beta_i}\Biggl\{
    \log{\prod_{i=1}^{N}\beta_i^{z_i}(1-\beta_i)^{1-z_i}}
	  - \log{p(\bm{z}\mid\Lambda; \bm{\alpha}^{(\tau)})}
    \Biggr\}
	= \frac{z_i}{\beta_i} - \frac{1-z_i}{1-\beta_i}.\nonumber
\end{align}
Thus, we solve Eq. \ref{eq:argmin_beta} as follows:
\begin{align}
	0&=\frac{\partial}{\partial \beta_i}\sum_{\bm{z}}
  \Biggl[
    \prod_{i=1}^{N}\beta_i^{z_i}(1-\beta_i)^{1-z_i}
	  \log{
      \frac{q'(\bm{z}\mid\Lambda; \bm{\beta})}
           {p(\bm{z}\mid\Lambda; \bm{\alpha}^{(\tau)})}
    }
  \Biggr]
  \nonumber\\
	&= \sum_{\bm{z}}\!
  \Biggl[q'(\bm{z}_{-i}\mid\Lambda, \bm{\beta})
  \Biggl\{
    z_i\!\Bigl(
      \frac{1\!-\!\beta_i}{\beta_i}
    \Bigr)^{1\!-\!z_i}\!
    -\!(1\!-\!z_i)\Bigl(
      \frac{\beta_i}{1\!-\!\beta_i}
    \Bigr)^{z_i}
  \Biggr\}
	\Biggl\{
    \log{\frac{q'(\bm{z}\!\mid\!\Lambda;\!\bm{\beta})}
              {p(\bm{z}\!\mid\!\Lambda;\!\bm{\alpha}^{(\tau)})}
        }
  \Biggr\}
  \nonumber\\
  &\qquad\qquad
	+\!q'(\bm{z}_{-i}\mid\Lambda, \bm{\beta})\beta_i^{z_i}(1\!-\!\beta_i)^{1\!-\!z_i}
	\Biggl\{
    \frac{z_i}{\beta_i}\!-\!\frac{1\!-\!z_i}{1\!-\!\beta_i}
  \Biggr\}
  \Biggr]
	\nonumber\\
	&=\sum_{\bm{z}}\!\Biggl[
    q'(\bm{z}_{-i}\mid\Lambda, \bm{\beta})\Biggl\{
      z_i\!\Bigl(
      \frac{1\!-\!\beta_i}{\beta_i}
      \Bigr)^{1\!-\!z_i}\!
    -\!(1\!-\!z_i)\Bigl(
      \frac{\beta_i}{1\!-\!\beta_i}
      \Bigr)^{z_i}
    \Biggr\}\Biggl\{
      \log{\frac{q'(\bm{z}\!\mid\!\Lambda;\!\bm{\beta})}
                {p(\bm{z}\!\mid\!\Lambda;\!\bm{\alpha}^{(\tau)})}
          }
      \Biggr\}
    \nonumber\\
    &\qquad\qquad
	  + q'(\bm{z}_{-i}\mid\Lambda, \bm{\beta})\Biggl\{
        z_i\!\Bigl(
        \frac{1\!-\!\beta_i}{\beta_i}
        \Bigr)^{1\!-\!z_i}\!
        -\!(1\!-\!z_i)\Bigl(
        \frac{\beta_i}{1\!-\!\beta_i}
        \Bigr)^{z_i}\!
      \Biggr\}
    \Biggr]
	\nonumber\\
	&=\sum_{\bm{z}}\!\Biggl[
    q'(\bm{z}_{-i}\mid\Lambda, \bm{\beta})\Biggl\{
      z_i\!\Bigl(
      \frac{1\!-\!\beta_i}{\beta_i}
      \Bigr)^{1\!-\!z_i}\!
      -\!(1\!-\!z_i)\Bigl(
      \frac{\beta_i}{1\!-\!\beta_i}
      \Bigr)^{z_i}
    \Biggr\}
	  \Biggl\{
      \log{\frac{q'(\bm{z}\!\mid\!\Lambda;\!\bm{\beta})}
                {p(\bm{z}\!\mid\!\Lambda;\!\bm{\alpha}^{(\tau)})}
          }\!+\!1
    \Biggr\}\Biggr].\nonumber
\end{align}
Therefore, the above equation becomes
\begin{align}
	&\sum_{\bm{z}} q'(\bm{z}_{-i}\mid\Lambda, \bm{\beta})z_i\Bigl(
      \frac{1-\beta_i}{\beta_i}
    \Bigr)^{1-z_i}
	  \Biggl\{
      \log{\frac{q'(\bm{z}\mid\Lambda;\bm{\beta})}
	              {p(\bm{z}\mid\Lambda;\bm{\alpha}^{(\tau)})}}+1
    \Biggr\}
	\nonumber\\
  &\qquad\qquad
  =\sum_{\bm{z}} q'(\bm{z}_{-i}\mid\Lambda, \bm{\beta})(1-z_i)\Bigl(
    \frac{\beta_i}{1-\beta_i}
    \Bigr)^{z_i}
	  \Biggl\{
      \log{\frac{q'(\bm{z}\mid\Lambda;\bm{\beta})}
	              {p(\bm{z}\mid\Lambda;\bm{\alpha}^{(\tau)})}}+1
    \Biggr\}
	\nonumber\\
	\Leftrightarrow&
	\sum_{\bm{z}_{-i}}q'(\bm{z}_{-i}\mid\Lambda, \bm{\beta})\Biggl\{
    \log{\frac{q'(\bm{z}_{-i}\mid\Lambda,z_i=1;\bm{\beta})}
	            {p(\bm{z}_{-i}\mid\Lambda,z_i=1;\bm{\alpha}^{(\tau)})}}+1
  \Biggr\}
  \nonumber\\
  &\qquad\qquad
  =\sum_{\bm{z}_{-i}} q'(\bm{z}_{-i}\mid\Lambda, \bm{\beta})\Biggl\{
    \log{\frac{q'(\bm{z}_{-i}\mid\Lambda,z_i=0;\bm{\beta})}
	            {p(\bm{z}_{-i}\mid\Lambda, z_i=0; \bm{\alpha}^{(\tau)})}}+1
  \Biggr\}
	\nonumber\\
	\Leftrightarrow&
	\sum_{\bm{z}_{-i}}q'(\bm{z}_{-i}\mid\Lambda, \bm{\beta})
  \log{\frac{q'(\bm{z}_{-i}\mid\Lambda,z_i=1;\bm{\beta})}
            {q'(\bm{z}_{-i}\mid\Lambda,z_i=0;\bm{\beta})
  }}
  \nonumber\\
  &\qquad\qquad
	= \sum_{\bm{z}_{-i}} q'(\bm{z}_{-i}\mid\Lambda, \bm{\beta})
  \log{\frac{p(\bm{z}_{-i}\mid\Lambda, z_i=1; \bm{\alpha}^{(\tau)})}
            {p(\bm{z}_{-i}\mid\Lambda, z_i=0; \bm{\alpha}^{(\tau)})}}
  \nonumber\\
	\Leftrightarrow&
	\log{\frac{\beta_i}{1-\beta_i}}
  = \sum_{\bm{z}_{-i}}q'(\bm{z}_{-i}\mid\Lambda, \bm{\beta})
    \log{\cfrac{p(\bm{z}_{-i}\mid\Lambda,z_i=1;\bm{\alpha}^{(\tau)})}
               {p(\bm{z}_{-i}\mid\Lambda,z_i=0;\bm{\alpha}^{(\tau)})}}.
\end{align}
Thus, we define the right-hand side in the above equation as $A^{(\tau)}$ and obtain
\begin{align}
	\beta_i^{(\tau+1)} \leftarrow \cfrac{1}{1+\exp{\bigl\{-A^{(\tau)}\bigr\}}}.
\end{align}

\subsection{Derivation of the projection from $\mathcal{D}'$ to $\mathcal{M}'$}
We consider $m$-projection from $\mathcal{D}'$ to $\mathcal{M}'$:
\begin{align}
	\bm{\alpha}^{(\tau+1)} = \argmin_{\bm \alpha} D_{\mathrm{KL}}[q'(\Lambda, \bm{z}; \bm{\beta}^{(\tau)}), p(\Lambda, \bm{z}; \bm{\alpha})].
	\label{eq:deriv8}
\end{align}
The KL divergence is explicitly given by
\begin{align}
	\lefteqn{
		D_{\mathrm{KL}}[q'(\Lambda, \bm{z}; \bm{\beta}^{(\tau)}), p(\Lambda, \bm{z}; \bm{\alpha})]
    }
  \qquad\nonumber\\
	&= \sum_{\Lambda}\sum_{\bm{z}} \Bigr\{q'(\Lambda, \bm{z}; \bm{\beta}^{(\tau)})\log{q'(\Lambda, \bm{z}; \bm{\beta}^{(\tau)})}\nonumber\\
	&\qquad\qquad\qquad- q'(\Lambda, \bm{z}; \bm{\beta}^{(\tau)})\log{p(\Lambda, \bm{z}; \bm{\alpha})}\Bigr\}.
	\label{eq:deriv9}
\end{align}
Therefore, the minimization in Eq. \ref{eq:deriv8} is equivalent to the following maximization:
\begin{align}
	\bm{\alpha}^{(\tau+1)}&=\argmax_{\bm \alpha}\sum_{\Lambda}\sum_{\bm{z}}q'(\Lambda, \bm{z}; \bm{\beta}^{(\tau)})\log{p(\Lambda, \bm{z}; \bm{\alpha})}
  \nonumber\\
	&=\argmax_{\bm \alpha}\sum_{\Lambda}\sum_{\bm{z}}q'(\Lambda)q'(\bm{z}\mid \Lambda; \bm{\beta}^{(\tau)})\log{p(\Lambda\mid \bm{z})p(\bm{z}; \bm{\alpha})}
  \nonumber\\
	&=\argmax_{\bm \alpha}\sum_{\Lambda}\sum_{\bm{z}}q'(\Lambda)q'(\bm{z}\mid \Lambda; \bm{\beta}^{(\tau)})\log{p(\bm{z}; \bm{\alpha})}
  \nonumber\\
	&=\argmax_{\bm \alpha}\sum_{\Lambda}q'(\Lambda)\sum_{\bm{z}}q'(\bm{z}\mid \Lambda; \bm{\beta}^{(\tau)})\log{\prod_{j=1}^N p(z_j; \alpha_j)}
  \nonumber\\
	&=\argmax_{\bm \alpha}\sum_{\bm{z}}q'(\bm{z}\mid \Lambda; \bm{\beta}^{(\tau)})\sum_{j=1}^N \log{p(z_j; \alpha_j)}.
	\label{eq:deriv10}
\end{align}
Partial differentiation of the right-hand side of Eq.\ref{eq:deriv10} with respect to $\alpha_i$ gives
\begin{align}
	0 &= \frac{\partial}{\partial\alpha_i} \Bigl\{\sum_{\bm z}q'(\bm{z}\mid \Lambda; \bm{\beta}^{(\tau)})\sum_{j=1}^N \log{p(z_j; \alpha_j)}\Bigr\}\nonumber\\
	&= \sum_{\bm z}q'(\bm{z}\mid \Lambda; \bm{\beta}^{(\tau)})\Bigl\{\frac{z_i}{\alpha_i} - \frac{1 - z_i}{1 - \alpha_i}\Bigr\}.
\end{align}
We can solve this equation as
\begin{align}
	&\sum_{\bm z}q'(\bm{z}\mid \Lambda; \bm{\beta}^{(\tau)})\frac{z_i}{\alpha_i} = \sum_{\bm z}q'(\bm{z}\mid \Lambda; \bm{\beta}^{(\tau)})\frac{1 - z_i}{1 - \alpha_i}\nonumber\\
	\Leftrightarrow& \sum_{\bm z}q'(\bm{z}\mid \Lambda; \bm{\beta}^{(\tau)})z_i(1 - \alpha_i) = \sum_{\bm z}q'(\bm{z}\mid \Lambda; \bm{\beta}^{(\tau)})(1 - z_i)\alpha_i\nonumber\\
	\Leftrightarrow& \sum_{\bm z}q'(\bm{z}\mid \Lambda; \bm{\beta}^{(\tau)})z_i = \sum_{\bm z}q'(\bm{z}\mid \Lambda; \bm{\beta}^{(\tau)})\alpha_i\nonumber\\
	\Leftrightarrow& \alpha_i = \sum_{\bm z}z_iq'(\bm{z}\mid \Lambda; \bm{\beta}^{(\tau)}),\nonumber
\end{align}
and obtain the following updated formula:
\begin{align}
	\alpha_i^{(\tau+1)} \leftarrow  \sum_{\bm z}z_iq'(\bm{z}\mid \Lambda; \bm{\beta}^{(\tau)}).
\end{align}

\section{Generation of network of 10,000 neurons}
We describe how to generate a network of 10,000 neurons similar to real-world data.
Here, we assume that these 10,000 neurons exist in the same cortical column.

\subsection{Location of neurons and generation of connections}
\begin{figure}[H]
  \centering
  \includegraphics[bb=0 0 417 423, width=.75\linewidth]{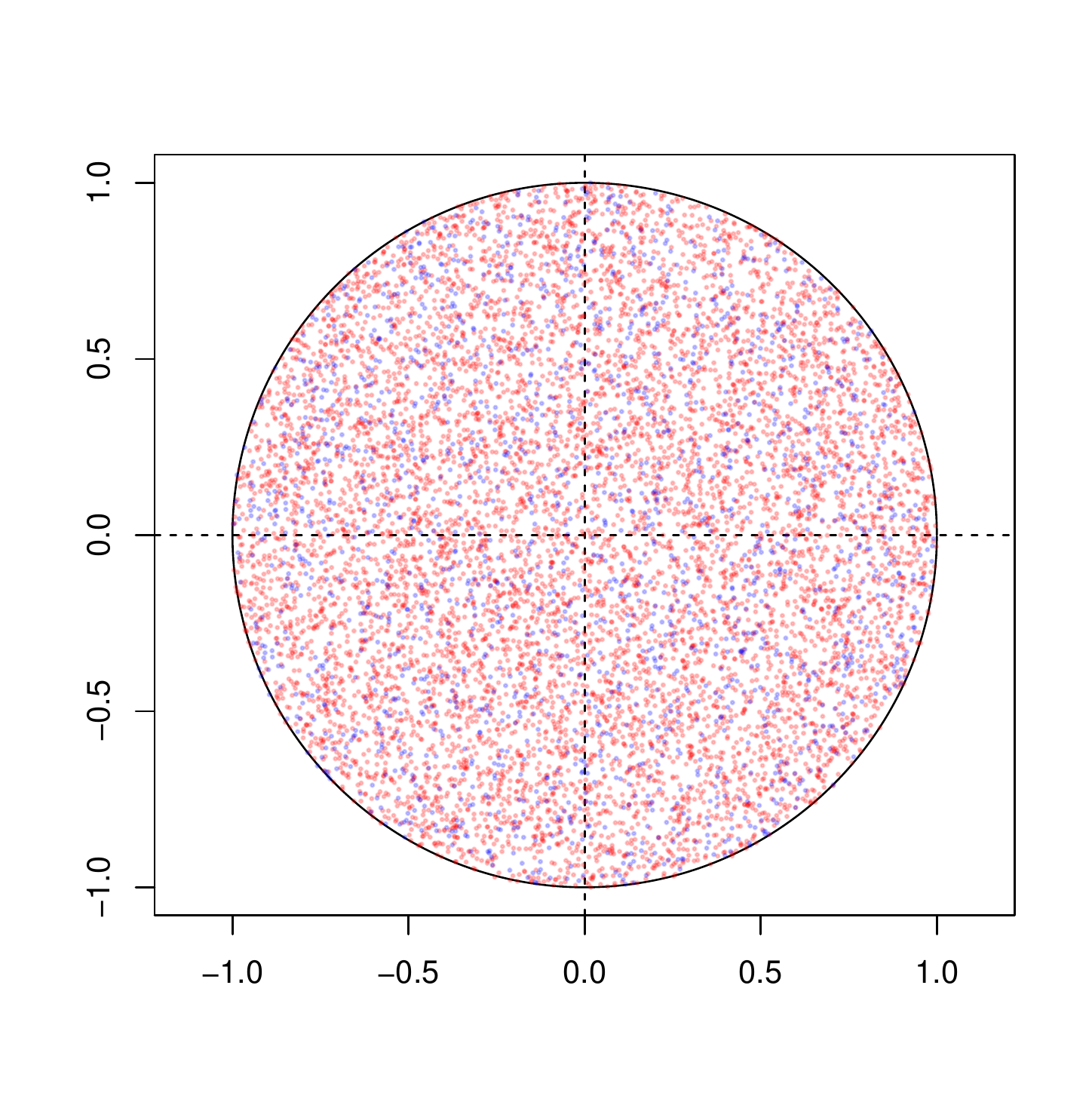}
  \caption{10,000 neurons located in a circular area having a radius of $1$[mm]. In this figure, the red points represent excitatory neurons and the blue points represent inhibitory neurons.}
  \label{fig:neurons_in_column}
\end{figure}

The 10,000 neurons are located randomly in a circular area having a radius of $1[{\rm mm}]$ with center coordinates $(0, 0)$, as shown in \Figref{fig:neurons_in_column}.
This circular area is regarded as a cortical column.
The coordinates of neuron $i$ are given by
\begin{align}
	x_i &= (x_i^{(1)}, x_i^{(2)}) = (r_i\cos{\theta_i}, r_i\sin{\theta_i}),
	\\
	r_i &= \sqrt{2u}, \quad u \sim \Bigl[0, \frac{1}{2}\Bigr],
	\\
	\theta_i &\sim [0, 2\pi].
\end{align}
We assume that the Euclidean distance $d_{ij}$ between directly connected neurons $i$ and $j$ follows a Gaussian distribution with mean $0$ and variance $\sigma_i^2$:
\begin{align}
	p^{(i)}_{j} &= \frac{f(d_{ij}\mid\sigma_i)}{\sum_{j\neq i}f(d_{ij}\mid\sigma_i)},
	\\
	f(x\mid\sigma_i) &= \sqrt{\frac{2}{\pi\sigma_i^2}}\exp{\Bigl(-\frac{x^2}{2\sigma_i^2}\Bigr)}.
\end{align}
The variance $\sigma_i$ depends on whether the neuron is excitatory or inhibitory.
It is known that inhibitory neurons connect more distantly located neurons than excitatory neurons.
On the basis of the settings in \citep{song2005highly}, $\sigma_i$ of an excitatory neuron is set as $75[{\rm \mu m}]$ and $\sigma_i$ of an inhibitory neuron is set as $250[{\rm \mu m}]$.

Then, we set the number of connections for each neuron.
Taking the circular shape of the area of interest into account, peripheral neurons may have fewer connections than neurons located at the center.
Therefore, the number of connections $M_i$ for peripheral neurons is defined by
\begin{align}
  M_i = \mathrm{round}(\gamma_iM),
\end{align}
where $\gamma_i$ is the ratio of the entire circle and the small circle around the peripheral neuron (for simplicity, the connection from the peripheral neuron is within this small circle), as shown in \Figref{fig:example_connections}.
$M$ is the maximum number of connections of neurons, i.e., following \citet{song2005highly}, we determined that around $10$\% of $400$ neurons sampled from a circular region with diameter $600[{\rm \mu m}]$ are mutually connected.
The function $\mathrm{round}(\cdot)$ is to round off to an integer value.
In this study, the range is a circular area with $r_i=3\sigma_i[{\rm \mu m}]$.
\begin{figure}[h]
  \centering
  \includegraphics[bb=0 0 612 792, width=.5\linewidth]{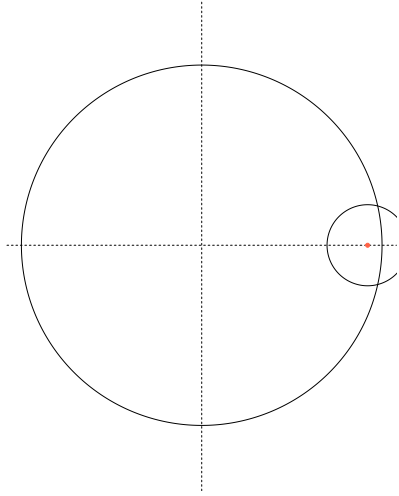}
  \caption{Circular area and range of connections for a peripheral neuron~(red point).}
  \label{fig:neuron_covered_area}
\end{figure}

We set the connection weights of each connection.
In this experiment, the excitatory weights are subject to a lognormal distribution based on \citep{song2005highly}~:
\begin{align}
  w_{ij} \sim \frac{1}{\sqrt{2\pi\sigma_w^2}w}
  \exp\Bigl\{
    -\frac{(\ln{w})^2}{2\sigma_w^2}
  \Bigr\},
\end{align}
where the parameter $\sigma_w$ is set to $1.5$ in this experiment.
The inhibitory weights are subject to uniform distribution $[-10, 0]$.

\subsection{Examples of synthesize networks}
\begin{figure}[H]
  \begin{minipage}{0.5\hsize}
    \centering
    \includegraphics[bb=0 0 460 460, trim=100 100 75 50, width=.8\linewidth]{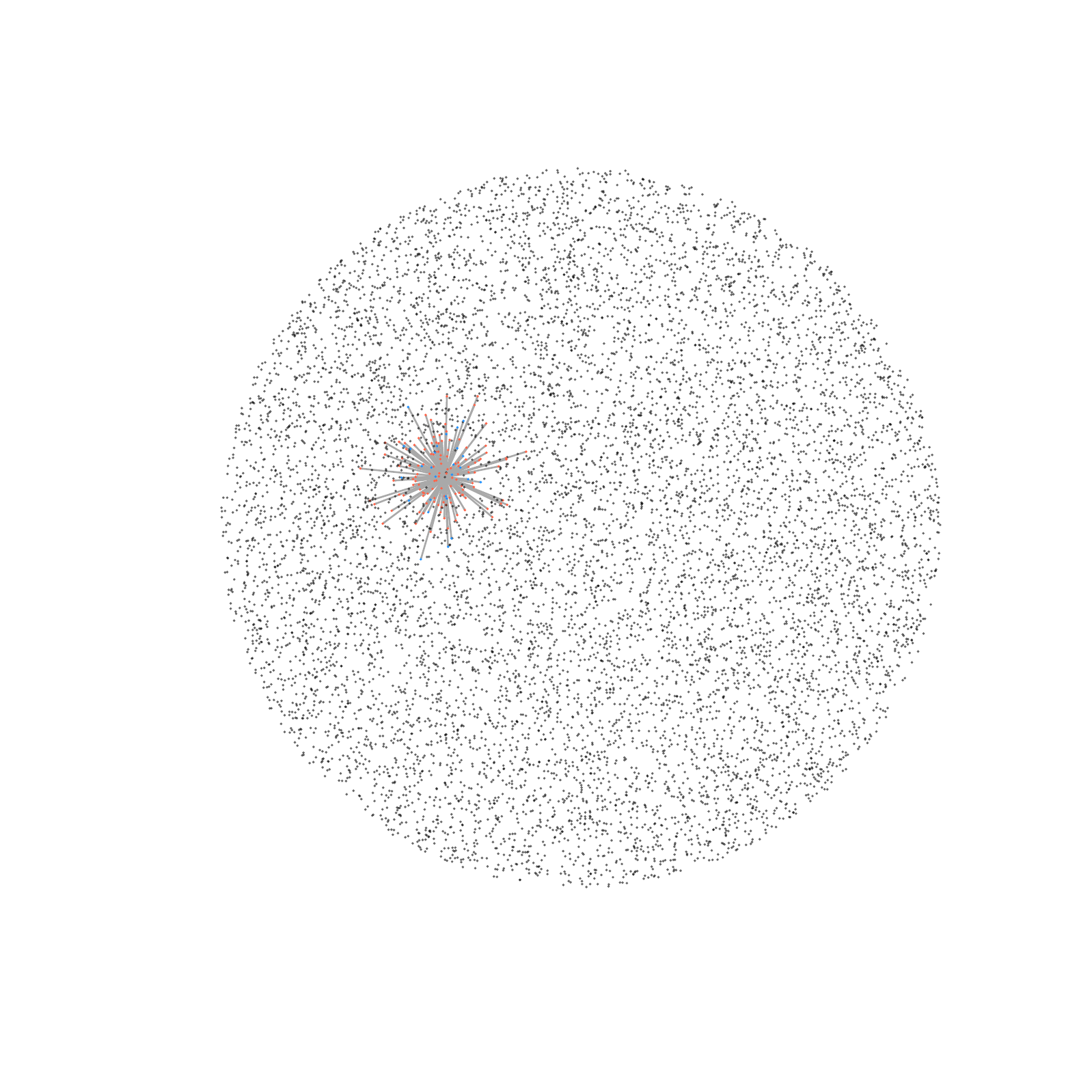}
  \end{minipage}
  \begin{minipage}{0.5\hsize}
    \centering
    \includegraphics[bb=0 0 460 460, trim=75 100 100 50, width=.8\linewidth]{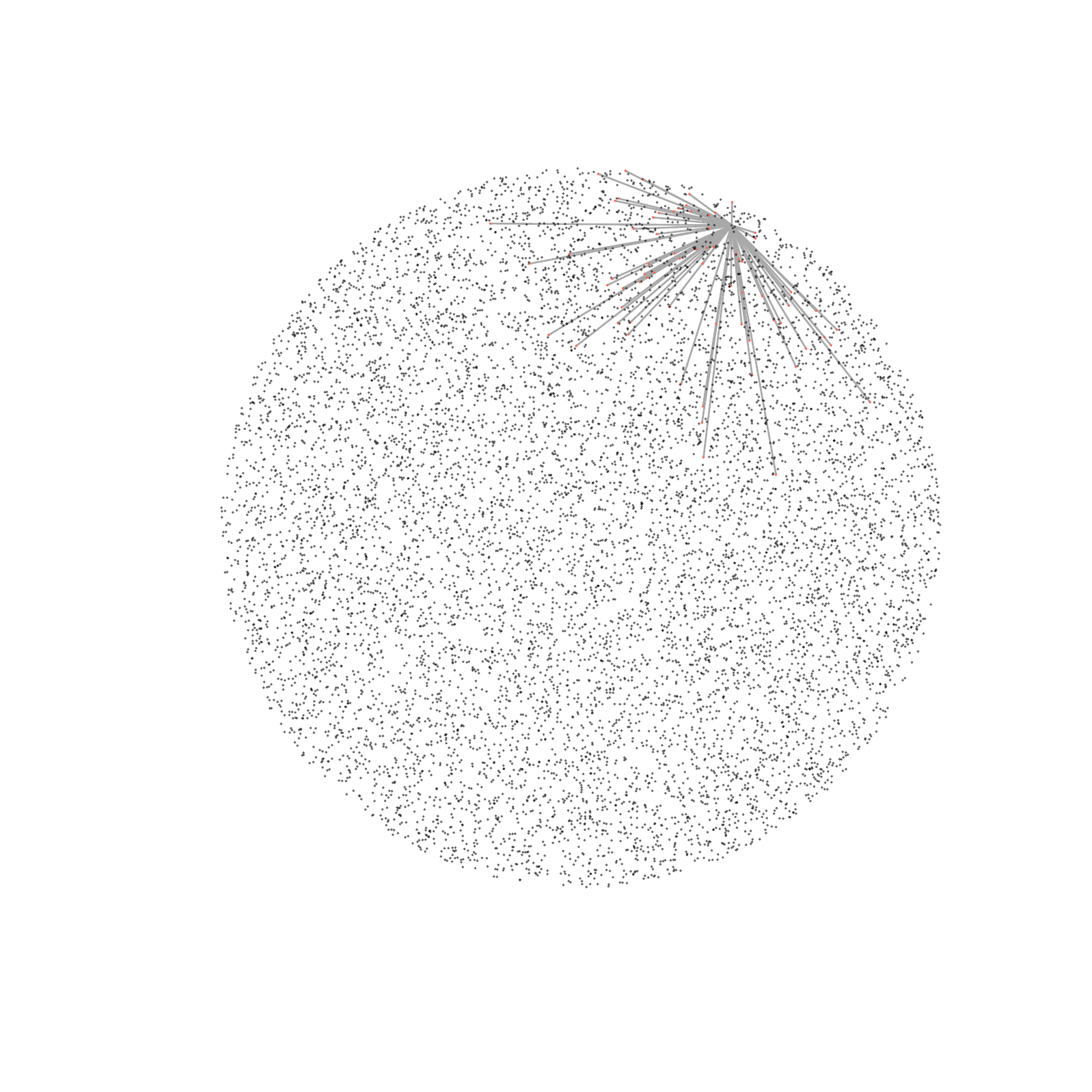}
  \end{minipage}
  \caption{Synthesized network of 10,000 neurons; connections of only one of them are visualized. Left: excitatory connections; Right: inhibitory connections.}
  \label{fig:example_connections}
\end{figure}

Examples of the synthesized networks are shown in \Figref{fig:example_connections}, in which connections from a single neuron~(excitatory on the left and inhibitory on the right) are visualized.



\begin{thebibliography}{34}
\providecommand{\natexlab}[1]{#1}
\providecommand{\url}[1]{\texttt{#1}}
\expandafter\ifx\csname urlstyle\endcsname\relax
  \providecommand{\doi}[1]{doi: #1}\else
  \providecommand{\doi}{doi: \begingroup \urlstyle{rm}\Url}\fi

\bibitem[Aertsen et~al.(1989)Aertsen, Gerstein, Habib, and
  Palm]{aertsen1989dynamics}
AM~Aertsen, GL~Gerstein, MK~Habib, and G~Palm.
\newblock Dynamics of neuronal firing correlation: modulation of "effective
  connectivity".
\newblock \emph{Journal of neurophysiology}, 61\penalty0 (5):\penalty0
  900--917, 1989.

\bibitem[Amari(1995)]{amari1995information}
Shun-Ichi Amari.
\newblock Information geometry of the em and em algorithms for neural networks.
\newblock \emph{Neural networks}, 8\penalty0 (9):\penalty0 1379--1408, 1995.

\bibitem[Arnold et~al.(2007)Arnold, Liu, and Abe]{arnold2007temporal}
Andrew Arnold, Yan Liu, and Naoki Abe.
\newblock Temporal causal modeling with graphical granger methods.
\newblock In \emph{Proceedings of the 13th ACM SIGKDD international conference
  on Knowledge discovery and data mining}, pages 66--75. ACM, 2007.

\bibitem[Banerjee et~al.(2008)Banerjee, Ghaoui, and
  d’Aspremont]{banerjee2008model}
Onureena Banerjee, Laurent~El Ghaoui, and Alexandre d’Aspremont.
\newblock Model selection through sparse maximum likelihood estimation for
  multivariate gaussian or binary data.
\newblock \emph{Journal of machine learning research}, 9\penalty0
  (Mar):\penalty0 485--516, 2008.

\bibitem[Barth{\'o} et~al.(2004)Barth{\'o}, Hirase, Monconduit, Zugaro, Harris,
  and Buzs{\'a}ki]{bartho2004characterization}
Peter Barth{\'o}, Hajime Hirase, Lena{\"\i}c Monconduit, Michael Zugaro,
  Kenneth~D Harris, and Gy{\"o}rgy Buzs{\'a}ki.
\newblock Characterization of neocortical principal cells and interneurons by
  network interactions and extracellular features.
\newblock \emph{Journal of neurophysiology}, 92\penalty0 (1):\penalty0
  600--608, 2004.

\bibitem[Bi and Poo(1998)]{bi1998synaptic}
Guo-qiang Bi and Mu-ming Poo.
\newblock Synaptic modifications in cultured hippocampal neurons: dependence on
  spike timing, synaptic strength, and postsynaptic cell type.
\newblock \emph{Journal of neuroscience}, 18\penalty0 (24):\penalty0
  10464--10472, 1998.

\bibitem[Brown et~al.(2004)Brown, Kass, and Mitra]{brown2004multiple}
Emery~N Brown, Robert~E Kass, and Partha~P Mitra.
\newblock Multiple neural spike train data analysis: state-of-the-art and
  future challenges.
\newblock \emph{Nature neuroscience}, 7\penalty0 (5):\penalty0 456--461, 2004.

\bibitem[Friedman et~al.(2008)Friedman, Hastie, and
  Tibshirani]{friedman2008sparse}
Jerome Friedman, Trevor Hastie, and Robert Tibshirani.
\newblock Sparse inverse covariance estimation with the graphical lasso.
\newblock \emph{Biostatistics}, 9\penalty0 (3):\penalty0 432--441, 2008.

\bibitem[Gerstein and Perkel(1969)]{gerstein1969simultaneously}
George~L Gerstein and Donald~H Perkel.
\newblock Simultaneously recorded trains of action potentials: analysis and
  functional interpretation.
\newblock \emph{Science}, 164\penalty0 (3881):\penalty0 828--830, 1969.

\bibitem[Gothard et~al.(1996)Gothard, Skaggs, and
  McNaughton]{gothard1996dynamics}
Katalin~M Gothard, William~E Skaggs, and Bruce~L McNaughton.
\newblock Dynamics of mismatch correction in the hippocampal ensemble code for
  space: interaction between path integration and environmental cues.
\newblock \emph{Journal of Neuroscience}, 16\penalty0 (24):\penalty0
  8027--8040, 1996.

\bibitem[Haury et~al.(2012)Haury, Mordelet, Vera-Licona, and
  Vert]{haury2012tigress}
Anne-Claire Haury, Fantine Mordelet, Paola Vera-Licona, and Jean-Philippe Vert.
\newblock Tigress: trustful inference of gene regulation using stability
  selection.
\newblock \emph{BMC systems biology}, 6\penalty0 (1):\penalty0 145, 2012.

\bibitem[Hino et~al.(2015)Hino, Takano, and Murata]{hino2015mmpp}
Hideitsu Hino, Ken Takano, and Noboru Murata.
\newblock mmpp: A package for calculating similarity and distance metrics for
  simple and marked temporal point processes.
\newblock \emph{R JOURNAL}, 7\penalty0 (2):\penalty0 237--248, 2015.

\bibitem[Hu et~al.(2015)Hu, Li, and Liang]{hu2015copula}
Meng Hu, Wu~Li, and Hualou Liang.
\newblock A copula-based granger causality measure for the analysis of neural
  spike train data.
\newblock \emph{IEEE/ACM Transactions on Computational Biology and
  Bioinformatics}, 2015.

\bibitem[Hyv\"arinen(1999)]{hyvarinen1999gaussian}
Aapo Hyv\"arinen.
\newblock Gaussian moments for noisy independent component analysis.
\newblock \emph{IEEE Signal Processing Letters}, 6\penalty0 (6):\penalty0
  145--147, 1999.

\bibitem[Ito and Tsuji(2000)]{ito2000model}
Hiroyuki Ito and Satoshi Tsuji.
\newblock Model dependence in quantification of spike interdependence by joint
  peri-stimulus time histogram.
\newblock \emph{Neural computation}, 12\penalty0 (1):\penalty0 195--217, 2000.

\bibitem[Izhikevich(2003)]{Izhikevich2003simple}
Eugene~M Izhikevich.
\newblock Simple model of spiking neurons.
\newblock \emph{IEEE Transactions on neural networks}, 14\penalty0
  (6):\penalty0 1569--1572, 2003.

\bibitem[Kim et~al.(2011)Kim, Putrino, Ghosh, and Brown]{kim2011granger}
Sanggyun Kim, David Putrino, Soumya Ghosh, and Emery~N Brown.
\newblock A granger causality measure for point process models of ensemble
  neural spiking activity.
\newblock \emph{PLoS Comput Biol}, 7\penalty0 (3):\penalty0 e1001110, 2011.

\bibitem[Kudrimoti et~al.(1999)Kudrimoti, Barnes, and
  McNaughton]{kudrimoti1999reactivation}
Hemant~S Kudrimoti, Carol~A Barnes, and Bruce~L McNaughton.
\newblock Reactivation of hippocampal cell assemblies: effects of behavioral
  state, experience, and eeg dynamics.
\newblock \emph{Journal of Neuroscience}, 19\penalty0 (10):\penalty0
  4090--4101, 1999.

\bibitem[Liu et~al.(2010)Liu, Roeder, and Wasserman]{liu2010stability}
Han Liu, Kathryn Roeder, and Larry Wasserman.
\newblock Stability approach to regularization selection (stars) for high
  dimensional graphical models.
\newblock In \emph{Advances in neural information processing systems}, pages
  1432--1440, 2010.

\bibitem[Markus et~al.(1994)Markus, Barnes, McNaughton, Gladden, and
  Skaggs]{markus1994spatial}
Etan~J Markus, Carol~A Barnes, Bruce~L McNaughton, Victoria~L Gladden, and
  William~E Skaggs.
\newblock Spatial information content and reliability of hippocampal ca1
  neurons: effects of visual input.
\newblock \emph{Hippocampus}, 4\penalty0 (4):\penalty0 410--421, 1994.

\bibitem[Nakahara and Amari(2002)]{nakahara2002information}
Hiroyuki Nakahara and Shun-ichi Amari.
\newblock Information-geometric measure for neural spikes.
\newblock \emph{Neural computation}, 14\penalty0 (10):\penalty0 2269--2316,
  2002.

\bibitem[Nie and Tatsuno(2012)]{nie2012information}
Yimin Nie and Masami Tatsuno.
\newblock Information-geometric measures for estimation of connection weight
  under correlated inputs.
\newblock \emph{Neural computation}, 24\penalty0 (12):\penalty0 3213--3245,
  2012.

\bibitem[Noda et~al.(2014)Noda, Hino, Tatsuno, Akaho, and
  Murata]{noda2014intrinsic}
Atsushi Noda, Hideitsu Hino, Masami Tatsuno, Shotaro Akaho, and Noboru Murata.
\newblock Intrinsic graph structure estimation using graph laplacian.
\newblock \emph{Neural computation}, 26\penalty0 (7):\penalty0 1455--1483,
  2014.

\bibitem[Perkel et~al.(1967)Perkel, Gerstein, and Moore]{perkel1967neuronal}
Donald~H Perkel, George~L Gerstein, and George~P Moore.
\newblock Neuronal spike trains and stochastic point processes.
\newblock 1967.

\bibitem[Quinn et~al.(2011)Quinn, Coleman, Kiyavash, and
  Hatsopoulos]{quinn2011estimating}
Christopher~J Quinn, Todd~P Coleman, Negar Kiyavash, and Nicholas~G
  Hatsopoulos.
\newblock Estimating the directed information to infer causal relationships in
  ensemble neural spike train recordings.
\newblock \emph{Journal of computational neuroscience}, 30\penalty0
  (1):\penalty0 17--44, 2011.

\bibitem[Ribeiro et~al.(2004)Ribeiro, Gervasoni, Soares, Zhou, Lin, Pantoja,
  Lavine, and Nicolelis]{ribeiro2004long}
Sidarta Ribeiro, Damien Gervasoni, Ernesto~S Soares, Yi~Zhou, Shih-Chieh Lin,
  Janaina Pantoja, Michael Lavine, and Miguel~AL Nicolelis.
\newblock Long-lasting novelty-induced neuronal reverberation during slow-wave
  sleep in multiple forebrain areas.
\newblock \emph{PLoS biology}, 2\penalty0 (1):\penalty0 e24, 2004.

\bibitem[Scheinberg and Rish(2009)]{scheinberg2009sinco}
Katya Scheinberg and Irina Rish.
\newblock {SINCO}-a greedy coordinate ascent method for sparse inverse
  covariance selection problem.
\newblock \emph{preprint}, 2009.

\bibitem[Shimazaki et~al.(2012)Shimazaki, Amari, Brown, and
  Gr{\"u}n]{shimazaki2012state}
Hideaki Shimazaki, Shun-ichi Amari, Emery~N Brown, and Sonja Gr{\"u}n.
\newblock State-space analysis of time-varying higher-order spike correlation
  for multiple neural spike train data.
\newblock \emph{PLoS Comput Biol}, 8\penalty0 (3):\penalty0 e1002385, 2012.

\bibitem[Song et~al.(2005)Song, Sj{\"o}str{\"o}m, Reigl, Nelson, and
  Chklovskii]{song2005highly}
Sen Song, Per~Jesper Sj{\"o}str{\"o}m, Markus Reigl, Sacha Nelson, and Dmitri~B
  Chklovskii.
\newblock Highly nonrandom features of synaptic connectivity in local cortical
  circuits.
\newblock \emph{PLoS biology}, 3\penalty0 (3):\penalty0 e68, 2005.

\bibitem[Takano et~al.(2015)Takano, Hino, Yoshikawa, and
  Murata]{takano2015patchworking}
Ken Takano, Hideitsu Hino, Yuki Yoshikawa, and Noboru Murata.
\newblock Patchworking multiple pairwise distances for learning with distance
  matrices.
\newblock In \emph{International Conference on Latent Variable Analysis and
  Signal Separation}, pages 287--294. Springer, 2015.

\bibitem[Tatsuno et~al.(2006)Tatsuno, Lipa, and
  McNaughton]{Tatsuno2006methodological}
Masami Tatsuno, Peter Lipa, and Bruce~L McNaughton.
\newblock Methodological considerations on the use of template matching to
  study long-lasting memory trace replay.
\newblock \emph{Journal of Neuroscience}, 26\penalty0 (42):\penalty0
  10727--10742, 2006.

\bibitem[Tatsuno et~al.(2009)Tatsuno, Fellous, and
  Amari]{tatsuno2009information}
Masami Tatsuno, Jean-Marc Fellous, and Shun-ichi Amari.
\newblock Information-geometric measures as robust estimators of connection
  strengths and external inputs.
\newblock \emph{Neural computation}, 21\penalty0 (8):\penalty0 2309--2335,
  2009.

\bibitem[Toyoizumi et~al.(2009)Toyoizumi, Rad, and Paninski]{toyoizumi2009mean}
Taro Toyoizumi, Kamiar~Rahnama Rad, and Liam Paninski.
\newblock Mean-field approximations for coupled populations of generalized
  linear model spiking neurons with markov refractoriness.
\newblock \emph{Neural computation}, 21\penalty0 (5):\penalty0 1203--1243,
  2009.

\bibitem[Wilson and McNaughton(1994)]{wilson1994reactivation}
Matthew~A Wilson and Bruce~L McNaughton.
\newblock Reactivation of hippocampal ensemble memories during sleep.
\newblock \emph{Science}, 265\penalty0 (5172):\penalty0 676--679, 1994.

\end{thebibliography}

\end{document}